\documentclass[sigchi]{acmart}

\AtBeginDocument{%
  }

\usepackage{enumitem}
\usepackage{CJKutf8}
\usepackage{graphicx}
\usepackage{multirow}
\usepackage{longtable}
\newcommand{\zxsucr}[1]{{\color{black}#1}}

\copyrightyear{2023}
\acmYear{2023}
\setcopyright{acmlicensed}\acmConference[CHI '23]{Proceedings of the 2023 CHI Conference on Human Factors in Computing Systems}{April 23--28, 2023}{Hamburg, Germany}
\acmBooktitle{Proceedings of the 2023 CHI Conference on Human Factors in Computing Systems (CHI '23), April 23--28, 2023, Hamburg, Germany}
\acmPrice{15.00}
\acmDOI{10.1145/3544548.3581465}
\acmISBN{978-1-4503-9421-5/23/04}

\begin{document}

\title{LipLearner: Customizable Silent Speech Interactions on Mobile Devices}

\author{Zixiong Su}
\affiliation{%
  \institution{The University of Tokyo}
  \city{Tokyo}
  \country{Japan}
}
\email{zxsu@g.ecc.u-tokyo.ac.jp}

\author{Shitao Fang}
\affiliation{%
  \institution{The University of Tokyo}
  \city{Tokyo}
  \country{Japan}
}
\email{fst@iis-lab.org}

\author{Jun Rekimoto}
\affiliation{%
  \institution{The University of Tokyo}
  \institution{Sony CSL Kyoto}
  \city{Kyoto}
  \country{Japan}
}
\email{rekimoto@acm.org}

\begin{abstract}
\zxsucr{Silent speech interface is a promising technology that enables private communications in natural language. However, previous approaches only support a small and inflexible vocabulary, which leads to limited expressiveness.} We leverage contrastive learning to learn efficient lipreading representations, enabling few-shot command customization with minimal user effort. Our model exhibits high robustness to different lighting, posture, and gesture conditions on an in-the-wild dataset. For 25-command classification, an F1-score of 0.8947 is achievable only using one shot, and its performance can be further boosted by adaptively learning from more data. This generalizability allowed us to develop a mobile silent speech interface empowered with on-device fine-tuning and visual keyword spotting. A user study demonstrated that with LipLearner, users could define their own commands with high reliability guaranteed by an online incremental learning scheme. \zxsucr{Subjective feedback indicated that our system provides essential functionalities for customizable silent speech interactions with high usability and learnability.}

\end{abstract}



\begin{teaserfigure}
  \includegraphics[width=\textwidth]{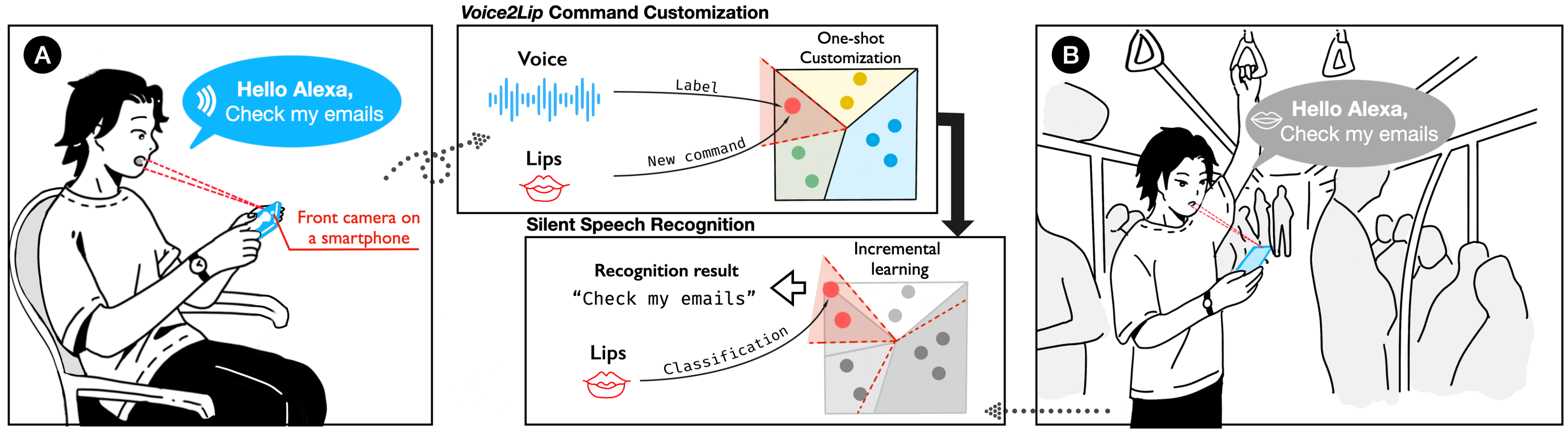}
  \caption{Example interaction of LipLearner. A) \textit{Voice2Lip} in-situ command registration. The user records a silent speech command by vocalizing it once, then LipLearner automatically learns to lip-read it with the text recognized from the voice signal as the label. B) The command then can be used \textit{without} vocalization, triggered by a silent keyword. LipLearner enables silent speech recognition which can be used in public settings (e.g., on the subway). Furthermore, it leverages incremental learning to proactively extend the model's knowledge when new samples become available.}
  \Description{}
  \label{fig:teaser}
\end{teaserfigure}

\maketitle

\section{Introduction}

Conversational agents are becoming increasingly integrated into our daily lives, serving as a fundamental element of ubiquitous computing and Internet-of-Things (IoT). They facilitate our approaches to edge devices by providing intuitive and efficient interactions, allowing people to communicate directly with devices in natural language. Thanks to the tremendous prevalence of smartphones, voice assistants~\cite{vui_us} have been unprecedentedly popular, giving users handy access to smartphone functionalities, smart home control, real-time information, and so forth. Despite the great convenience offered by voice input, there are three major limitations hampering its usability in practice. Voice User Interfaces (VUIs) 1) is not a preferred option in public settings due to the risk of privacy and security problems, and people may feel awkward talking to a smartphone in front of others~\cite{milanesi2016voice}, 2) relies on accurate speech recognition, which can be difficult when the environment is noisy, and 3) is not available for people with speech disorders. 

To tackle the privacy and social acceptance issues in VUIs, Silent Speech Interface (SSI) has emerged as a promising alternative that exploits non-acoustic signals to enable speech recognition without voice. SSI provides seamless and confidential interactions in various situations, especially in those where voice interaction is inappropriate or unavailable. Recent research on SSI has proposed to use various sensing methods such as Electromyography (EMG)~\cite{kapur2018alterego,meltzner2018development}, ultrasound imaging~\cite{kimura2019sottovoce}, capacitive sensing~\cite{li2019tongueboard} and video camera (lipreading) ~\cite{lipinteract,kimura2020tielent,pandey2021liptype} to track the movement of speech articulators and decode silent speech.


We focus on the last method, which is also known as lipreading, and augment it with few-shot learning to enable customizable silent speech commands on smartphones. Compared to other approaches, lipreading has minimal device requirements but provides rich information with high temporal and spatial resolution. Nowadays, smartphones have become the most popular devices and most of them are equipped with high-quality digital cameras. Therefore, implementing lipreading systems on smartphones further pushes forward the convenience and lowers the bar of silent speech input. On the other hand, as the primary input method on smartphones, touch gestures can be cumbersome when only single-handed input is available. For such situations, silent speech has been proven stable and efficient as a supplementary input modality~\cite{lipinteract}.

However, there are three major challenges to building expressive lipreading systems in practice. First, the data collection process should minimize the effort for new users to get started with. However, previous approaches to SSIs, not limited to lipreading-based approaches, adopt a train-from-scratch model that requires collecting hundreds of samples from real users~\cite{lipinteract, su2021gaze+, kimura2020tielent, speechin}, leading to excessive mental and physical user burden. Second, such data collected intensively in controlled laboratory environments causes a biased model, which can be sensitive to even minor changes in factors such as lighting, face orientations, and postures, yet there is little discussion on the model's ability to generalize to unseen environments. Third, the model training process is time-consuming and requires high-end GPUs, but they are not always accessible to users for many reasons (requiring internet connection, high computing cost, privacy concerns of uploading face data to the cloud, etc.). Adding new commands is even more difficult, because it requires collecting new data as well as re-training the model from scratch. As a result, only a limited number of pre-defined commands are available, and the rich interaction space in silent speech is still waiting to be mined.


In this research, our goal is to liberate the expensiveness of lipreading as well as reduce the user burden in the data collection process. We propose a few-shot lipreading framework that enables in-situ silent speech command customization. We set off by pre-training a lipreading encoder model using a contrastive learning objective, which learns efficient and robust visual speech representations from public datasets in a semi-supervised manner. We then employ a simple linear classifier, which can be trained in a few seconds, to transfer the model to unseen users and words using a few-shot learning strategy. Hence, the user can freely define any phrase in any language, or even non-verbal lip gestures, as a silent speech command by providing at least one sample. We further minimize the user effort of enrolling new commands by introducing \textit{Voice2Lip}, a multimodal command registration technique that automatically learns lipreading from voice input in a one-shot manner. To register a new command, the user just says it aloud, and then our system will learn the lip movements using the text recognized from voice signals as labels.

To ensure the applicability of our method in the real world, we performed a model test under diverse conditions that covered a broad range of daily scenarios, including different lighting conditions, body postures, and holding gestures. The results show that our few-shot customization framework could achieve an F1-score of 0.8947 in unseen conditions with only one shot, significantly outperforming the conventional user-dependent approach though the latter used four times more training data.
We then built a mobile application called LipLearner on a commodity smartphone.


To empower LipLearner with reliable hands-free activation, we propose a visual keyword spotting method that detects the user-defined keyword from lip movements. Most previous lipreading interfaces use the mouth opening degree (MOD)~\cite{su2021gaze+, lipinteract} as the only cue to trigger silent speech input, which is prone to misactivation. For example, the system can easily respond to the user's unintentional behavior, such as smiling or talking to others. 
Our lipreading model encodes lip movements into embedding vectors, which can be used to identify the keyword from continuous input by computing the cosine similarity. This function can also be customized and initialized with only a few positive samples (i.e., no negative samples required). 
Moreover, we introduced an online incremental learning scheme to allow users to continually improve the performance of the model by providing new samples during interaction. With the efficient lip embeddings, it is trivial to fine-tune the model by only updating the liner classifier, which significantly reduces the computing resource demand and thus allows all real-time customizations and recognitions to be performed on-device for privacy preservation.
Our quantitative user study results show that LipLearner could recognize 30 commands (20 of which were user-dependent) with a one-shot accuracy of 81.7\%. The performance improved gradually while more samples were provided by the user; finally, 98.8\% accuracy was achieved with five samples per command. Subjective feedback indicates that our system was easy to use and learn, and the human-AI interaction experience was enjoyed by many participants. \zxsucr{We have made our machine learning scripts, models, and the source code of LipLearner (iOS App) publicly available at \url{https://github.com/rkmtlab/LipLearner} to facilitate further work.}



In summary, this paper makes four key contributions:

1. A semi-supervised lipreading encoder that exploits public datasets to learn efficient visual speech representations and a few-shot silent speech customization framework to support novel commands with a small number of samples.

2. A model test demonstrating our method works robustly in a variety of environment and interaction factors, namely lighting conditions, body postures, and holding gestures. 

3. A mobile application that provides real-time and customizable silent speech interactions, empowered by a visual keyword spotting method for hands-free activation and an online incremental learning scheme for extendable vocabulary and performance.


4. A comprehensive user study that evaluated the system's real-world performance and usability with customizable commands.
%

\begin{table*}[t]
\centering
\resizebox{\textwidth}{!}{
\begin{tabular}{c|llcclll}
\toprule
\textbf{Paradigm}                   & \textbf{Research}            & \textbf{Device}             & \textbf{Vocabulary}             & \textbf{Samples} & \textbf{Accuracy} & \textbf{Activation}               & \textbf{Language}                   \\ \hline
\multirow{4}{*}{User-dependent model}  & Kimura et al. 2020~\cite{kimura2020tielent}                  & Wearable camera             & 15                              & 40              & 94\%              & Offline                           & English                             \\
                                       & Chen 2020 et al.~\cite{cface}                  & Wearable camera             & 8                               & 10              & 84.70\%           & Offline                    & English                             \\
                                       & Su et al. 2021~\cite{su2021gaze+}               & Fixed camera                & 27 (6\textsuperscript{\textdagger})                         & 18              & 91.63\%           & MOD                               & English                             \\
                                       & Zhang et al. 2021~\cite{speechin}                 & Wearable IR camera                  & 54/44                           & 24              & 90.5\%/91.6\%     & Offline                    & English/Chinese                     \\ \hline
\multirow{4}{*}{Off-the-shelf model}   & Sun et al. 2018~\cite{lipinteract}            & Smartphone                  & 44 (6-10\textsuperscript{\textdagger})                     & -             & 95.40\%           & MOD                               & Chinese                             \\
                                       & Saitoh and Kubowaka 2019~\cite{Saitoh2019LiP25wWL}                    & Smartphone                  & 25                              & -             & 73.40\%           & Manual                            & Japanese                            \\
                                       & Laxmi and Sabbir 2021~\cite{pandey2021liptype}                  & Smartphone                  & 51\textsuperscript{\textdagger\textdagger}                            & -             & WER 40.9\%        & Manual                            & English                             \\
                                       & Zhang et al. 2021~\cite{speechin}                 & Wearable IR camera                  & 54/44                           & -              & 54.4\%/61.2\%     & Offline                    & English/Chinese \\ \hline
\multirow{3}{*}{\textbf{Few-shot transfer learning}} & \textbf{LipLearner (1-shot)} & \multirow{3}{*}{\textbf{Smartphone}} & \multirow{3}{*}{\textbf{30\textsuperscript{\textdagger\textdagger\textdagger}}} & \textbf{1}      & \textbf{81.7\%}         & \multirow{3}{*}{\textbf{Keyword}} & \multirow{3}{*}{\textbf{Arbitrary}} \\
                                       & \textbf{LipLearner (3-shot)} &                             &                                 & \textbf{3}      & \textbf{96.5\%}         &                                   &                                     \\
                                       & \textbf{LipLearner (5-shot)} &                             &                                 & \textbf{5}      & \textbf{98.8\%}         &                                   &                                     \\

\bottomrule
\end{tabular}
}
\caption{Machine learning (ML) paradigms and their specifications in recent lipreading interfaces. The sample column shows the number of training samples the user needs to record for each command. \textsuperscript{\textdagger} The actual vocabulary size depends on the context. \textsuperscript{\textdagger\textdagger} The vocabulary is word-level. \textsuperscript{\textdagger\textdagger\textdagger} The vocabulary is custom (defined by each user). While some research only conducted offline experiments or asked the user to trigger the recognizer manually, LipLearner offers online keyword activation and recognition and is evaluated via a live user study.}
\label{tab: relatedwork}
\end{table*}

\section{Related Work}

In this section, we overview related literature in the domains of silent speech interfaces and relevant machine learning techniques.

\subsection{Silent Speech Interface}
Silent speech interfaces have been a research topic of vast research interest for decades, aiming to provide confidential and seamless communications. Similar to VUIs, SSIs allow users to converse with computers in natural language, which provides expressive commands without requiring them to remember complicated actions or gestures. Existing SSIs are characterized by what kind of sensing methods and biosignals are used, such as tracking the movement of speech articulators using electromagnetic articulography (EMA) ~\cite{fagan2008development, schonle1987electromagnetic, gonzalez2016silent}, vocal tract imaging using ultrasound imaging~\cite{hueber2010development, kimura2019sottovoce}, capturing subtle sounds produced by non-audible murmur (NAM)~\cite{toda2005nam, toda2009voice, toda2012statistical} and ingressive speech~\cite{fukumoto2018silentvoice}, placing capacitive sensors inside the mouth~\cite{li2019tongueboard, kimura2022silentspeller}, and capturing facial electrical activity using electromyography (sEMG)~\cite{wand2011session, kapur2018alterego}. In the field of Brain-Computer Interfaces (BCI), researchers seek to decode human speech directly from the electrical activity of the brain, where the approaches can be categorized into invasive systems implanted in the cerebral cortex using electrocorticography (ECoG) ~\cite{angrick2019speech, rabbani2019potential} and non-invasive systems attached to the scalp using Electroencephalogram (EEG)~\cite{porbadnigk2009eeg, guenther2009wireless, herff2015brain}. 

The most related literature to this work is lipreading-based SSIs. Lipreading is a technology that utilizes a camera to visually capture movement around the mouth and interpret speech from the image sequence. HCI researchers have proposed to use devices such as smartphones~\cite{lipinteract, pandey2021liptype} and wearable cameras~\cite{kimura2020tielent, cface, speechin} to provide mobile silent speech interaction, as well as multimodal approaches such as using silent speech to facilitate eye-gaze-based selection~\cite{su2021gaze+}. 

The challenges in lipreading stem from the inherent ambiguity of lip movements. The number of distinguishable visemes (i.e., minimum visual speech units) is usually considered to be 10-16~\cite{visser1999classifying, ezzat1998miketalk, ezzat2000visual}, which is much less than the number of phonemes. Researchers have proposed using ultrasound imaging to track movements of the oral cavity and tongue as a complementary method for lipreading~\cite{kimura2022ssr7000,kimura2020end,ji2018updating}. While this multimodal approach could significantly improve the performance of silent speech recognition, ultrasound imaging devices are cumbersome and impractical for mobile interactions. In contrast, our few-shot lipreading framework enables customizable and reliable silent speech interactions using only a commodity mobile phone.

\subsection{Machine Learning Approaches to Lipreading Interfaces}

Recent work in the deep learning field has shown the effectiveness of using deep neural networks (DNN) for lipreading ~\cite{feng2020learn, martinez2020lipreading, petridis2018end}, while the machine learning paradigms used to build such a model can have a significant impact on its performance. 

As shown in Table~\ref{tab: relatedwork}, we broadly divided previous lipreading interfaces into two categories: 1) user-dependent models, which collect data from each user and train individual models from scratch, and 2) off-the-shelf-models, which leverage either public datasets or pre-collected data to enable user-independent recognition. User-dependent models offer better performance because they have obtained knowledge from actual users. However, this method imposes a huge burden on new users, making it a difficult trade-off between the vocabulary and the amount of training data. Off-the-shelf models are available immediately without requiring new data. Nonetheless, building a model that can generalize to unseen users remains a huge challenge, as conventional methods either have a small vocabulary~\cite{lipinteract} or limited accuracy~\cite{Saitoh2019LiP25wWL, pandey2021liptype}. One workaround is to use a context-dependent vocabulary to improve accuracy, but it also limits the number of available commands at a time~\cite{su2021gaze+, lipinteract}. Furthermore, a common issue in both user-dependent models and off-the-shelf models is that the commands are pre-defined by the researchers. Making changes to the command set requires tremendous training data and computing resources, which are not accessible to users. Additionally, there is a lack of a practical activating method to initiate silent speech input. Previous methods such as offline segmentation~\cite{kimura2020tielent, cface, speechin} or trigger buttons~\cite{Saitoh2019LiP25wWL, pandey2021liptype} are not feasible for hands-free real-time interactions, and MOD-based methods can be vulnerable to misactivations~\cite{su2021gaze+, lipinteract}. We propose a novel few-shot transfer learning paradigm to enable customizable silent speech commands. LipLearner can achieve promising accuracy with a small amount of training samples, thus making it possible for the user to add arbitrary new commands. The few-shot paradigm also opens the door for \textit{visual} keyword spotting, which enables using silent speech to wake up devices. 

The generalizability of our model benefits from the contrastive pre-training pipeline. The weak supervision thereof makes the model more flexible when transferring to a new data domain, outperforming supervised approaches~\cite{chen2021empirical}. Recent research on using contrastive learning for lipreading has been focusing on learning from unlabeled audio-visual data ~\cite{crosslip, callip}. Although the abundant information of audio signals provoked an array of inspiring work, such as synthesizing speech from lips~\cite{9747427, prajwal2020learning}), localizing sounds in video frames~\cite{arandjelovic2018objects, senocak2018learning}, and separating speech signals~\cite{gao2021visualvoice}, it could bring unnecessary complexity to silent speech recognition. Our work differs from the previous ones in that we leverage a more straightforward approach that only utilizes the visual modality to obtain efficient representations with rich semantic information.

\subsection{Few-shot Transfer Learning in Human-Computer Interaction} 

Few-shot learning (FSL) is a deep learning paradigm, where a model is first pre-trained on large datasets and then fine-tuned using a few new samples to generalize to unseen data distributions. Instead of training the entire model from scratch each time, FSL can incrementally obtain new knowledge by only partially updating the model. HCI researchers have been applying FSL to tasks, such as sound recognition~\cite{protosound, listenlearner} and human activity recognition~\cite{cpc_har}, enabling in-situ model fine-tuning in the real world.
One of the most relevant literature is few-shot gesture recognition~\cite{enable_gesture}, as gestures and lip movements are both time series human motion signals. This work utilizes the IMU signals from a smartwatch to enable users to add custom gestures with a few samples. However, the model was pre-trained in a supervised manner, which could limit the model's generalizability: although the system applied data augmentation (which was performed on a laptop) to obtain more data for fine-tuning, the 1-shot accuracy was only 55.3\% in 12-gesture classification. Our approach leverages semi-supervised learning to learn more efficient representations, achieving high accuracy with a more lightweight architecture that can be deployed on a smartphone.

\begin{figure*}[t]
    \centering
    \makebox[0pt]{\includegraphics[width=\textwidth]{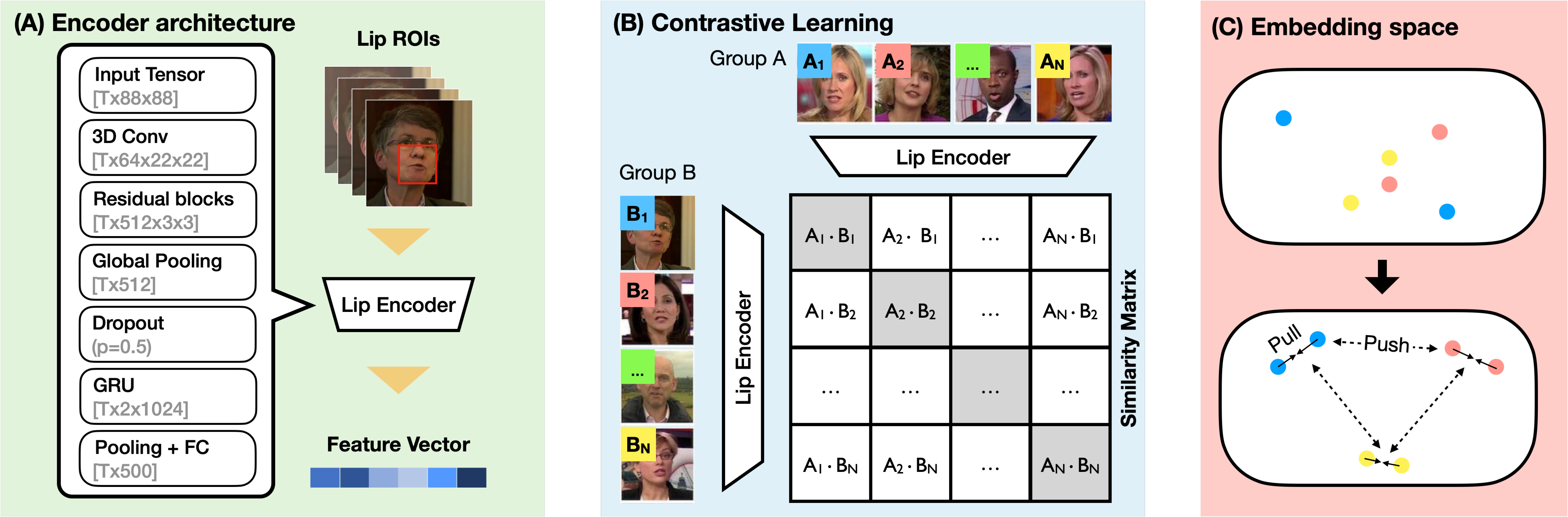}}
    \caption{The pre-learning pipeline. (A) We use a 3D CNN encoder to extract a low-dimensional feature vector from lip images. (B) The contrastive objective maximizes the similarities between utterances of the same words (diagonal elements in the similarity matrix) while minimizing similarities between utterances of different words (non-diagonal elements in the similarity matrix). The subscript numbers indicate the class indexes. (C) The learned embedding space.} 
    \label{fig:network}
\end{figure*}

\section{Contrastive Pre-training}

To overcome the limitation of vocabulary as well as minimize the user burden in the data collection process, we leverage contrastive learning to exploit knowledge from public datasets. In this section, we elaborate on the methods and techniques we used in this pre-training process, including the large-scale lipreading dataset, the neural network architecture, and the training details. The pre-trained lipreading encoder is the cornerstone of our few-shot customization framework.
\subsection{Pre-training Dataset and Preprocessing}

We use a public large-scale lipreading dataset, LRW~\cite{chung2016lip}, which comprises video segments extracted from the BBC news, to pre-train a robust feature extractor for few-shot lipreading. The dataset consists of up to 1000 utterances of 500 different words, spoken by hundreds of different speaker, thus providing rich utterances and face patterns. The speaker's face is cropped with the mouth centered using a facial landmark detection algorithm~\cite{kazemi2014one} provided by the Dlib Library~\cite{dlib09}. The dataset also covers diverse recording conditions, such as lighting, background, and camera perspective, which is expected to enhance the performance of model in real-world settings.

Nonetheless, there are still discrepancies between the data distributions of LRW and mobile silent speech scenarios. For instance, most videos in LRW were captured with fixed or stabilized cameras from a third person point of view. While in our scenarios, handheld devices, such as smartphones, inevitably lead to shaking videos, and their wide-angle lens can cause barrel distortion. Additionally, all LRW videos are sampled to 29 frames at 25fps (1.16 seconds), which can make the model sensitive to variations in video duration. To fill this gap, we apply several data augmentations to generate more data simulating smartphone videos, namely random crop, random frame drop, random shaking, and random barrel distortion. Finally, the frames were converted to grayscale and resized to 88 (H) $\times$ 88 (W) pixels.

\subsection{Model Architecture}
\label{sec:model}

We adopt an encoder model based on the architecture proposed in~\cite{feng2020learn}, which has achieved a state-of-the-art level performance in lipreading classification tasks. As shown in Figure~\ref{fig:network}, the neural network first extracts both spatial and temporal information using ResNet-18 with a 3D convolutional architecture. After a global pooling layer, the output is reshaped into $T \times 512$ (T denotes time). We then apply the same word boundary technique described in~\cite{feng2020learn}, which appends a binary vector to indicate the duration of the keyword. Finally, the feature is processed sequentially using a bidirectional Gated Recurrent Unit (GRU) followed by an average pooling and a fully connected layer, outputting a 500-dimensional feature vector.

\subsection{Contrastive Learning Pipeline}

Conventional supervised learning uses labeled data to learn to classify the inputs into known classes. The vocabulary of LRW consists of 500 individual words, which is, however, biased and far from allowing natural communications with smart assistants (e.g., "Question" and "Questions" take up 2 classes, but there are no words such as "What" for interrogative expression which is essential for a conversational interface). To overcome this limitation, we leverage contrastive learning, in which the objective is to learn an embedding space where similar samples are close to each other while dissimilar ones are far apart. Thus, we can use the model to find the most similar command when given samples, even if the samples belong to previously unseen classes.

In our implementation, we use the CLIP objective~\cite{radford2021learning} to let the model only learn the similarity between samples without remembering the exact label. As shown in Figure~\ref{fig:network} (B), we randomly select one sample from each of $ N $ ($N$ = batch size) classes as group $ A $, and then select another $N$ samples from the same classes as $ B $. Next, the samples are encoded into embeddings, and a cosine similarity matrix is calculated among the embeddings, scaled by a temperature parameter $\tau$:

\[S_{i,j} = sim(A_i, B_j) / \tau \]

Here, we use the same $\tau$ of 0.07 as CLIP. The cosine similarity sim(,) is measured by the dot product \zxsucr{of two L2-normalized embedding vectors $A_i$ and $B_j$, where $ i,j \in [0,N]$ denote the class indexes.} Note that unlike CLIP used different encoders for text and image data, our data only has the visual channel. Therefore, the encoders for the two data groups share the same weights. Diagonal values in the matrix are similarities between embeddings from the same class, while non-diagonal values are those between different classes. The model contrastively learns from the positive $N$ pairs and the negative $N^2 - N$ pairs using the InfoNCE loss~\cite{van2018representation}, which averages the cross entropy loss of group $A$ and group $B$.

\[ \mathcal{L} = -\frac{1}{2N}(\sum\nolimits_{i=1}^{N}\log{\frac{e^{S_{i,i}}}{\sum_j{e^{S_{i,j}}}}} + \sum\nolimits_{j=1}^{N}\log{\frac{e^{S_{j,j}}}{\sum_i{e^{S_{i,j}}}}})\]

\subsection{Training details}
The training starts from pre-trained weights provided by Feng et al.~\cite{feng2020learn}. We use a ReduceLROnPlateau scheduler with an initial learning rate of $3 \times 10^{-4}$, which is reduced by a factor of 0.5 once the validation loss stagnates for 40 epochs. The training loss converged after 500 epochs, taking around 34 hours across 2 NVIDIA GeForce RTX 2080 Ti GPUs. We save the model with the least loss on the validation set for our system. 

\section{Data Collection for Model Test}
\label{sec:data_collection}
%
There are many variables that could affect the performance of the lip encoder model. Particularly, we seek to analyze the model's robustness against challenges such as different environment configurations and user behaviors. To this end, we set off by collecting an in-the-wild dataset that covers various practical settings that simulate mobile interaction scenarios.

\subsection{Command Set}

\begin{figure*}[t]
    \centering
    \makebox[0pt]{\includegraphics[width=0.9\textwidth]{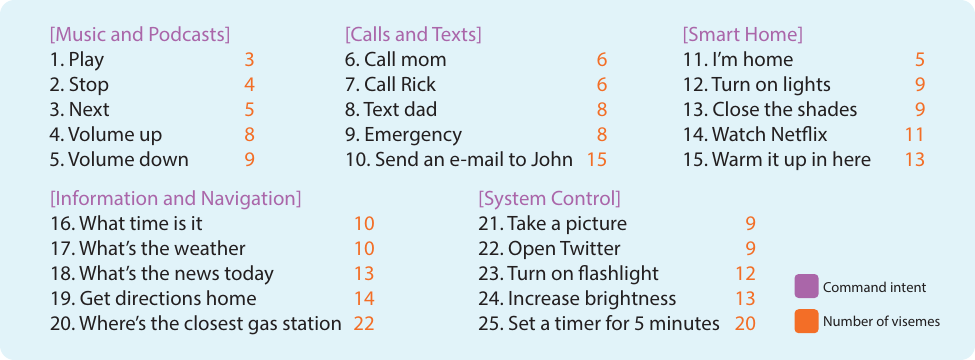}}
    \caption{Command set used for the model test.}
    \label{fig:command_set}
\end{figure*}

First of all, we designed a 25-sentence corpus for silent speech interaction (see Figure~\ref{fig:command_set}). This command set is intended to contextualize a scenario where people interact with a conversational assistant to operate the smartphone, control smart home devices, or find information. The phrases are partially selected from the most popular Alexa commands according to a recent research~\cite{sciuto2018hey}, and the rest are from Apple's official guide to Siri~\cite{applesiri}. We include both concise commands and casual expressions, covering all kinds of visemes and various lengths (3-22 visemes, average length $10.08 \pm 4.47$; we first translate the words to phonemes by referring to the CMU Pronouncing Dictionary~\cite{CMUdict} and then map the phonemes into visemes using Lee and Yook's approach~\cite{lee2002audio}). Therefore, this corpus is also phonetically well-balanced and suitable for testing the model's capability.

\subsection{Recording Conditions}

\begin{figure*}[t]
    \centering
    \makebox[0pt]{\includegraphics[width=0.9\textwidth]{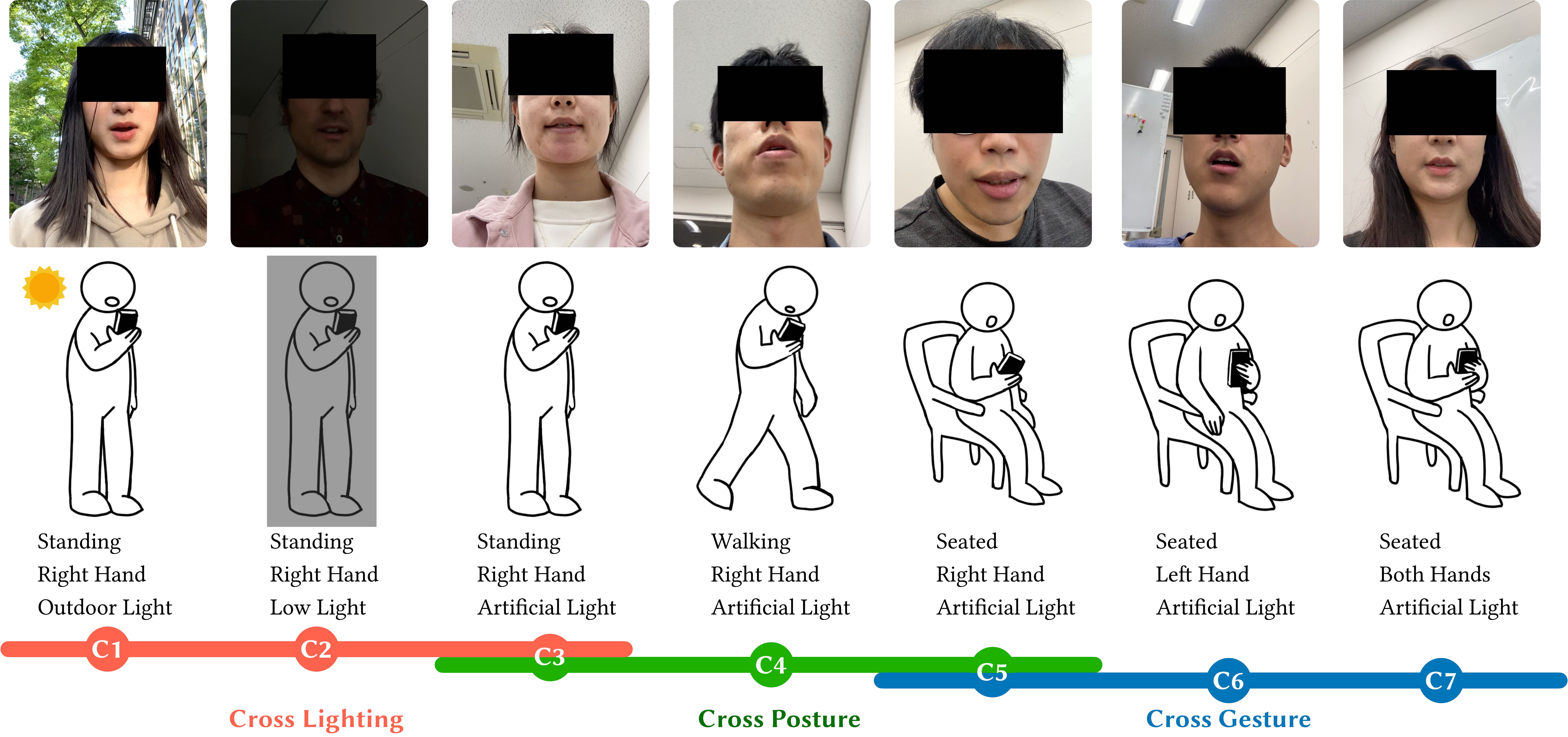}} \caption{Illustration of seven conditions during data collection and their corresponding captured views. Selected frames are processed for privacy protection. The recording conditions are intended for cross-lighting, cross-posture, and cross-gesture tests.}
    \label{fig:datacollection}
\end{figure*}

A mobile interface should provide stable performance across different conditions. Especially, we consider that there are three key factors, namely lighting, posture, and grasp gesture, that pose challenges to silent speech recognition. In this section, we elaborate on the various recording conditions contained in the dataset.

\subsubsection{Lighting}

We change the recording locations and time of day to achieve different luminance levels. Further investigations show that those daily scenarios can have a wide light intensity range.
 
\begin{itemize}[noitemsep,topsep=0pt]
    \item \textbf{Outdoor Daylight}: outdoor environment on sunny afternoons (1:00 PM - 3:00 PM). 
    \item \textbf{Low Light}: laboratory environment on later afternoons (3:00 PM - 5:00 PM), simulated by partially blocking the natural light.
    \item \textbf{Artificial Light}: laboratory environment with good lighting provided, natural light is blocked. 

\end{itemize}

\subsubsection{Posture}
Participants were asked to record while seated, standing, or walking. Different postures could cause different levels of shaking, leading to blurry videos and varying face positions.

\begin{itemize}[noitemsep,topsep=0pt]
    \item \textbf{Standing}: participants are asked to stand in place.
    \item \textbf{Walking}: participants are asked to record while walking along a straight line.
    \item \textbf{Seated}: participants are seated on a chair with their hands placed on the armrest.

\end{itemize}

\subsubsection{Grasp Gesture}

Participants were asked to hold the smartphone with their right hand, left hand, or both hands. Different grasp gestures result in significant differences in the face orientation relative to the camera.  

\begin{itemize}[noitemsep,topsep=0pt]
    \item \textbf{Right Hand}: the smartphone is held with the user's right hand.
    \item \textbf{Left Hand}: the smartphone is held with the user's left hand.
    \item \textbf{Both Hands}: the smartphone is held with the user's both hands.
\end{itemize}

\subsection{Procedure}

We recruited 11 participants (4 females and 7 males) from the local university, all right-handed. Note that to distinguish from the user study section, participants in this section are identified as \textit{speakers} (S1-S11). We used iPhone 11 and iPhone 13 Pro for video recording. All videos are saved in a MOV format with 1080 (H) $\times$ 1920 (W) pixels at 30 fps. In the collection process, the user is asked to press the record button at the bottom of the screen and then subvocalize the command prompt shown on the top. During the speech, the user needs to keep pressing the recording button and release as soon as they finish speaking to indicate the beginning and the end of the recording. Next, the subsequent command will be prompted. To avoid errors caused by unfamiliarity, we ask the user to read each of the commands at least once before collecting. If the user did not read the command correctly or fluently, they can use the rollback button at the bottom-right corner to record the last command again. 

The data collection was approved by the university's Institutional Review Board (IRB), and all participants have filled out an IRB-approved consent form. All participants completed seven collection sessions, each of which is under a condition that is a combination of the three key factors (see Figure~\ref{fig:datacollection}). For each session, participants were tasked to repeat each of the 25 commands five times. Between the sessions, participants were allowed to take a one-minute break. This procedure took around 40 minutes, and we compensated the participant 1050JPY for their time. In total, 11 participants $\times$ 7 sessions $\times$ 25 commands $\times$ 5 repetitions $=$ 9625 data points were collected.


\section{Customization Pipeline and Model Performance}

This section presents the few-shot tuning pipeline used to recognize novel silent speech commands with very few samples. Furthermore, we performed a comprehensive test to show that our approach is robust to a wide range of environment configurations.

\subsection{Pre-processing and Data Visualization} 

We extracted the mouth region from our study data using the MediaPipe face detector~\cite{lugaresi2019mediapipe} to identify the face landmarks. For each frame, we cropped a square region of interest (ROI) with the mouth centered according to the landmarks, which describes the location of the mouth. The ROI was converted to a grayscale image and then resized to 88 (H) $\times$ 88 (W) pixels, which follows the same pre-processing procedure as the LRW dataset. With the pre-trained lip encoder model, we embedded the ROI into a 500-dimension feature vector as a semantic representation of the silent speech command.

\begin{figure*}[t]
    \centering
    \makebox[0pt]{\includegraphics[width=15cm]{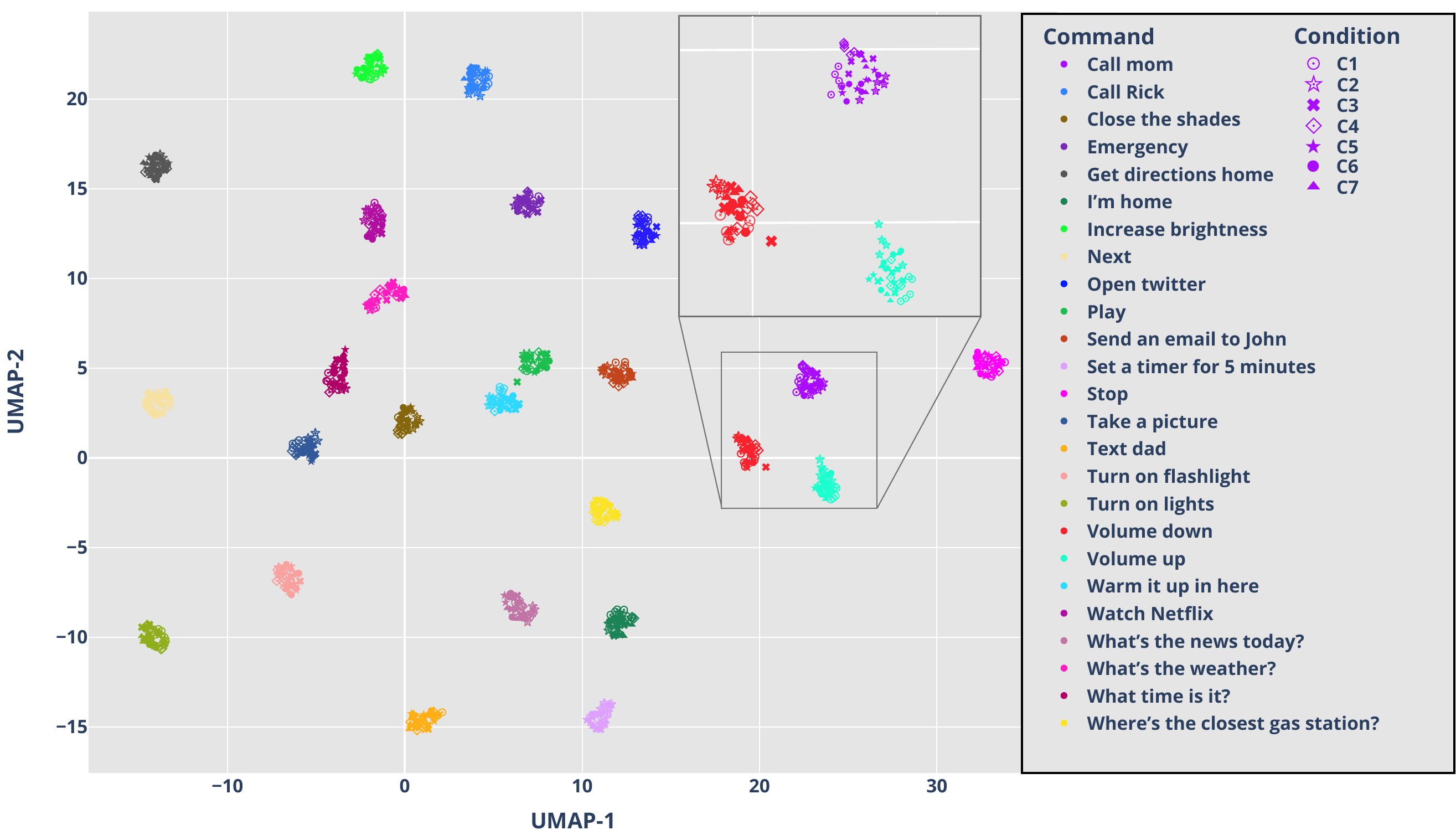}}
    \caption{2D UMAP Visualization of the feature embedding space with data from S10 as an example. Commands and conditions are depicted in colors and symbols, respectively. The zoom-in area shows that the data distributions of the same command from different conditions are mostly overlapped, suggesting that our visual speech representation is robust to environment factors.}
    \label{fig:umap}
\end{figure*}

To better understand how the feature vectors are distributed, we use the uniform manifold approximation and projection (UMAP) to visualize a subset of data obtained from a single speaker (S10) in a 2D space. UMAP is an unsupervised dimensionality reduction technique that clusters the data points without accounting for the labels in the transformation. As shown in Figure~\ref{fig:umap}, there are 25 distinct clusters corresponding to the 25 commands in the command set, which are linearly separable. In addition, our model exhibits a good generalization ability. For example, when zooming into two of the clusters (\textit{"Call mom"}, and \textit{"Volume up"}), it was unlikely to separate the data by the recording condition. Moreover, the distance between different conditions was much larger than that between different commands. 
Similar observations were also found in other users' collected data, which supports our assumption that the encoder model has learned efficient semantic representation that can be generalized to unseen speakers and phrases.

\subsection{Few-shot Fine-tuning Architecture}

Instead of directly computing the similarity, we used a simple linear logistic regression classifier, which is shown sufficient to achieve high accuracy with a very small amount of training samples~\cite{chen2021empirical, chen2019closer}, to learn novel commands. Logistic regression is adept at fitting linearly separable data, which is suitable for the highly abstracted features extracted by the encoder model. In the fine-tuning stage, we freeze the weights of the encoder model and only train the linear classifier, thus making it trivial to perform in-situ command customization on mobile devices. Note that the linear classifier is user-dependent and trained on each user's data to maximize accuracy. 

To better understand the capability and limitations of the silent speech representations, we conducted comprehensive experiments to test the model's performance in different dimensions. 

\subsection{Experiment 1: Effect of number of commands and number of shots}
\label{sec:moreshots}

\begin{figure*}[t]
    \centering
    \makebox[0pt]{\includegraphics[width=\textwidth]{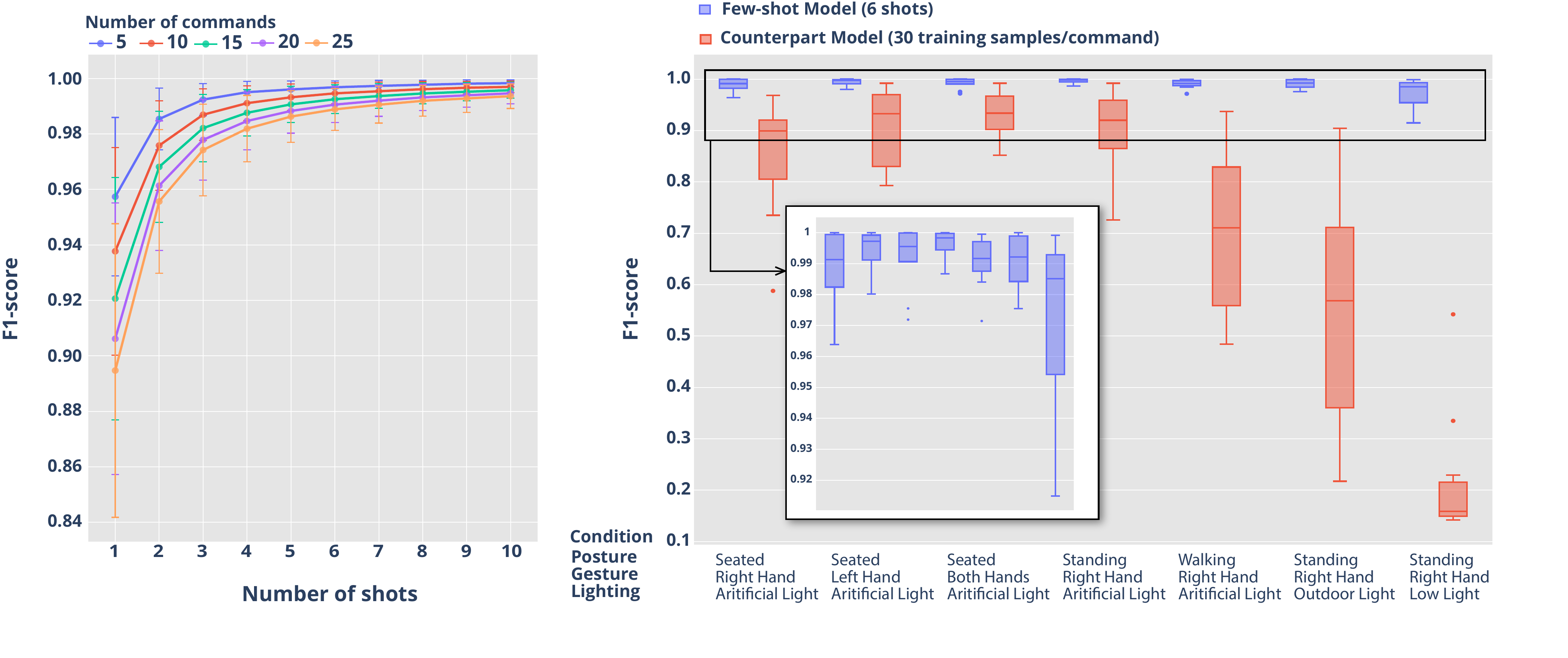}}
    \caption{Model test results in F1 measure. Left: Effect of the number of commands and the number of shots. Right: Generalization ability test.}
    \label{fig:model_test}
\end{figure*}

Our in-situ customization framework allows the user to enroll new commands or provide new samples for existing commands anytime and anywhere. We used our dataset to simulate this dynamic process and investigated how the number of commands and shots would affect recognition accuracy. In this session, we first randomly selected $M$ commands ($M \in \{5, 10, 15, 20, 25\}$). The last two shots from all conditions are selected as test data. We then trained the model with $N$ ($N \in [1..10] $) shot(s) randomly selected from the remaining data, which can belong to different conditions. 
Since there are too many possible combinations of data selection, we repeated the test 1000 times to simulate that training data is collected over various conditions in daily use. As illustrated in Figure~\ref{fig:model_test}, The model's performance improved rapidly as the number of shots increased. In 5-command classification, the F1-score was $0.9574 \pm 0.0286$ with only one shot and became $0.9924 \pm 0.0058$ with three shots of each command. Compared to other input modalities (e.g., gesture, eye gaze), one of the most important advantages of speech is its expressiveness. Therefore, supporting more commands is crucial to providing better silent speech interactions. 
The result showed that although more commands led to slight performance degradation, the model still obtains a one-shot F1-score of $0.8947 \pm 0.0530$ when classifying 25 commands and an F1-score of $0.9819 \pm 0.0120$ was achieved with four shots. The standard deviation was also reduced when the number of shots was increased, indicating that more training samples can improve the model's robustness. Thus, the proposed method is promising for recognizing a large number of silent speech commands, and the model's knowledge can be extended by collecting more data in real use.

\begin{table*}[t]
\centering
\begin{tabular}{ccccccc}

\toprule
 & & \multicolumn{5}{c}{Number of shots} \\
\hline
                   & Left-out condition     &1  &2  &3  &4  &5     \\
\hline
Cross Lighting      &\textbf{Average}   &0.8954	&0.9227	&0.9391	&0.9463	&0.9504     \\
                  & Artificial Light    &0.9079	&0.9344	&0.9509	&0.9587	&0.9629		       \\
                  & Outdoor Daylight    &0.8876	&0.9125	&0.9283	&0.9347	&0.9391         \\
                  & Low Light           &0.8906	&0.9212	&0.9382	&0.9454	&0.9493		  \\
\hline           
Cross Posture   &\textbf{Average}       &0.9189	&0.9436	&0.9595	&0.9665	&0.9702     \\
                 & Standing             &0.9291	&0.9504	&0.9637	&0.9697	&0.9717         \\
                 & Walking              &0.9183	&0.9431	&0.9601	&0.9669	&0.9717		    \\
                 & Seated               &0.9093	&0.9374	&0.9546	&0.9629	&0.9674         \\
                 
\hline
Cross Gesture    &\textbf{Average}      &0.9332	&0.9555	&0.9680	&0.9746	&0.9780   \\
                   & Right Hand         &0.9162	&0.9425	&0.9568	&0.9646	&0.9689		 \\
                   & Left Hand          &0.9456	&0.9632	&0.9739	&0.9797	&0.9823 \\
                    & Both Hand         &0.9377	&0.9609	&0.9733	&0.9796	&0.9828 \\

\bottomrule
\end{tabular}
\caption{Cross-condition model performance in F1 measure.}
\label{tab:cross_test}
\end{table*}

\subsection{Experiment 2: Generalization ability}

A common scenario is that the recording setting is significantly different from where the user actually uses it. The model can learn these differences by asking the user to provide samples in every possible condition, which however leads to user burden implications. We believe that our approach can be generalized to completely unseen conditions without having such training data. First, we ran a leave-one-condition-out test by training the classifier on data from six conditions and testing on data from the one remaining condition. For each training condition, we randomly selected only one sample from each class, forming a 6-shot training dataset. This test was repeated 100 times with random seeds. 
The box plot in Figure~\ref{fig:model_test} illustrates the distribution of the F1-scores for the 11 participants. To compare with the predominant approach, which trains the model from scratch with considerable data collected from real users, we built a counterpart model that had the same architecture as our encoder but was trained in a supervised fashion. The counterpart model was trained on all data obtained from the training conditions (i.e., 6 conditions $\times$ 5 repetitions $ =  $ 30 training samples per command), and it corresponds to the user-dependent train-from-scratch model in previous literature. Overall, our method achieved an F1-score of $0.9895 \pm 0.0078$ (averaged over conditions), surpassing the counterpart model's score of $ 0.7147 \pm 0.2576$. This result shows that our method provides significantly higher recognition accuracy and is more robust to unseen environments. In addition, the counterpart model exhibited worse performance especially in the last three conditions: walking posture (F1-score $= 0.6930$), outdoor light (F1-score $= 0.5510$), and low light (F1-score $= 0.2134$). This indicates that the accuracy of the conventional train-from-scratch method can be most severely affected by shaking videos and varying illuminations. To investigate our method's capability to cope with this problem, we further conducted cross-condition experiments with control variables in the following sections.

\subsection{Cross-condition Performance}

People use smartphones in different places and at different times, leading to varying lighting conditions that can significantly affect the video's quality. For example, insufficient lighting requires longer exposure time and higher sensor sensitivity, which can result in blurry images with noise. In contrast, bright sunlight can cause overexposed images that lacked highlight details. We select the data recorded under conditions C1, C2, and C3, corresponding to outdoor daylight, low light, and artificial light, respectively, while the keeping posture and grasp gesture are fixed to standing and right-hand holding. A cross-lighting test was conducted by training the classifier under two conditions and testing under the other condition.

The gesture of holding a smartphone depends on personal habits and the usage scenario. As a result, the camera angle relative to the face can vary in a wide range, causing different distortion effects in the image. We ran a cross-gesture test across conditions C3, C4, and C5, corresponding to standing, walking, and seated postures, respectively. While the gesture and lighting were set to right hand and artificial light.

Similarly, posture is also a vital factor in mobile lipreading, taking a video while walking leads to frequent camera angle changes and shaking videos with blurry frames. The cross-posture test was performed across conditions C5, C6, and C7, where the user was seated under artificial lights but with different gestures, namely right hand, left hand, and both hands.

All cross-condition tests were repeated 1000 times to mitigate the randomness of data selection. The results are shown in Table~\ref{tab:cross_test} with all conditions showing a similar trend: the more shots, the better performance. We also find that the cross-lighting condition was more challenging, as its 3-shot average F1-score was 0.9391, which was notably lower than the cross-posture and cross-gesture conditions (F1-score 0.9595 and 0.9680). Overall, we conclude our framework still shows high and robust performance even in unseen conditions, which is promising for real-world applications.

\section{LipLearner: Customizable and Learnable Silent Speech Assistant}

To investigate the usability of our silent speech customization method, we implemented LipLearner, a mobile application for in-situ customizable silent speech interaction with online few-shot learning. In this section, we elaborate on the implementation details of the application, including visual keyword spotting (KWS), online learning scheme, and interface design.

\begin{figure*}[t]
    \centering
    \makebox[0pt]{\includegraphics[width=0.95\textwidth]{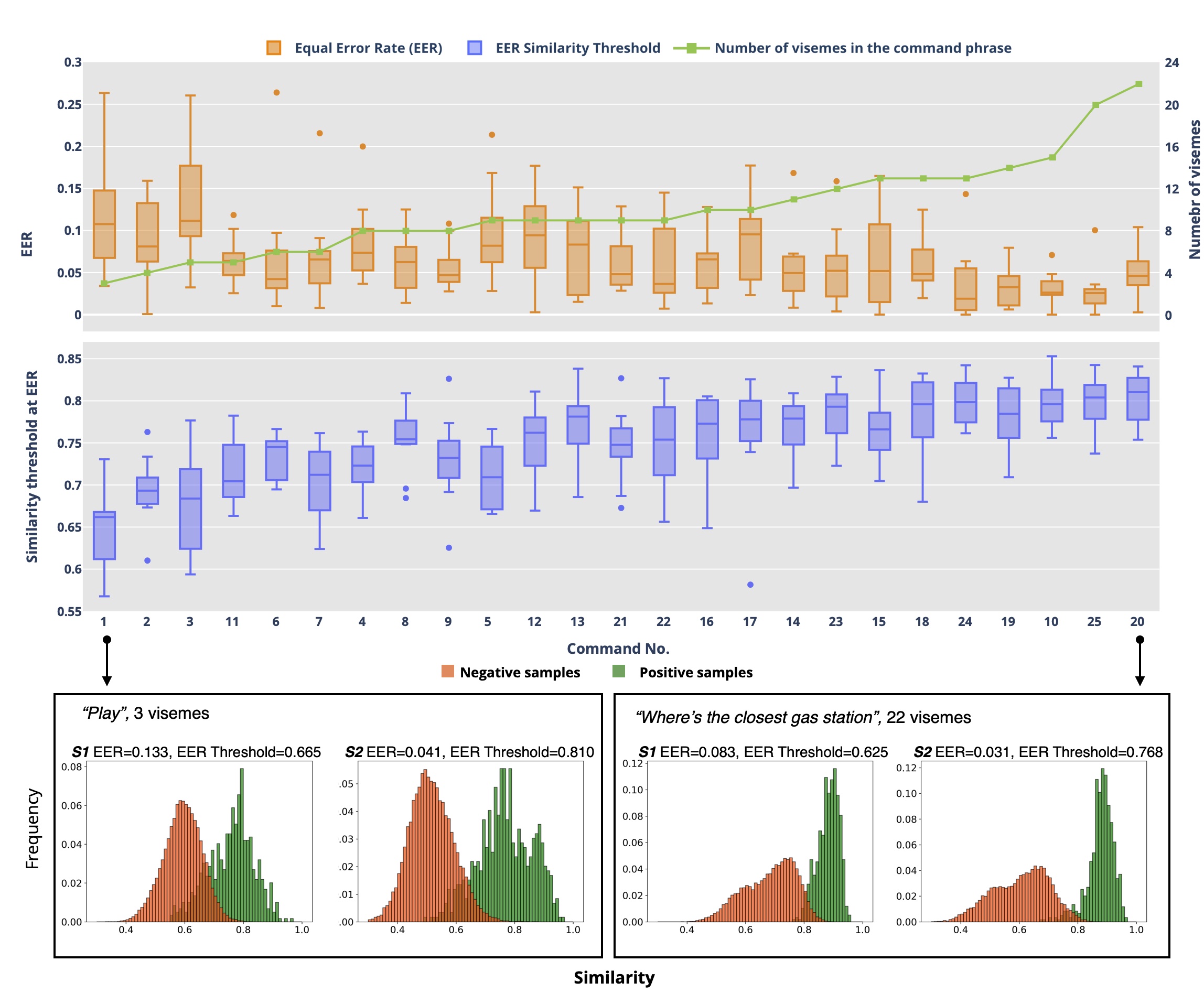}}
    \caption{Top: Per-command EER and EER similarity threshold. We find that commands consisting of more visemes have higher EER and EER thresholds. Note that for better visualization, the commands are sorted by the length in viseme. Bottom: an example illustration of the positive and negative distributions of command "Play" and "Where's the closest gas station," subvocalized by S1 and S2.}
    \label{fig:eer}
\end{figure*}

\begin{figure*}[t]
    \centering
    \makebox[0pt]{\includegraphics[width=16cm]{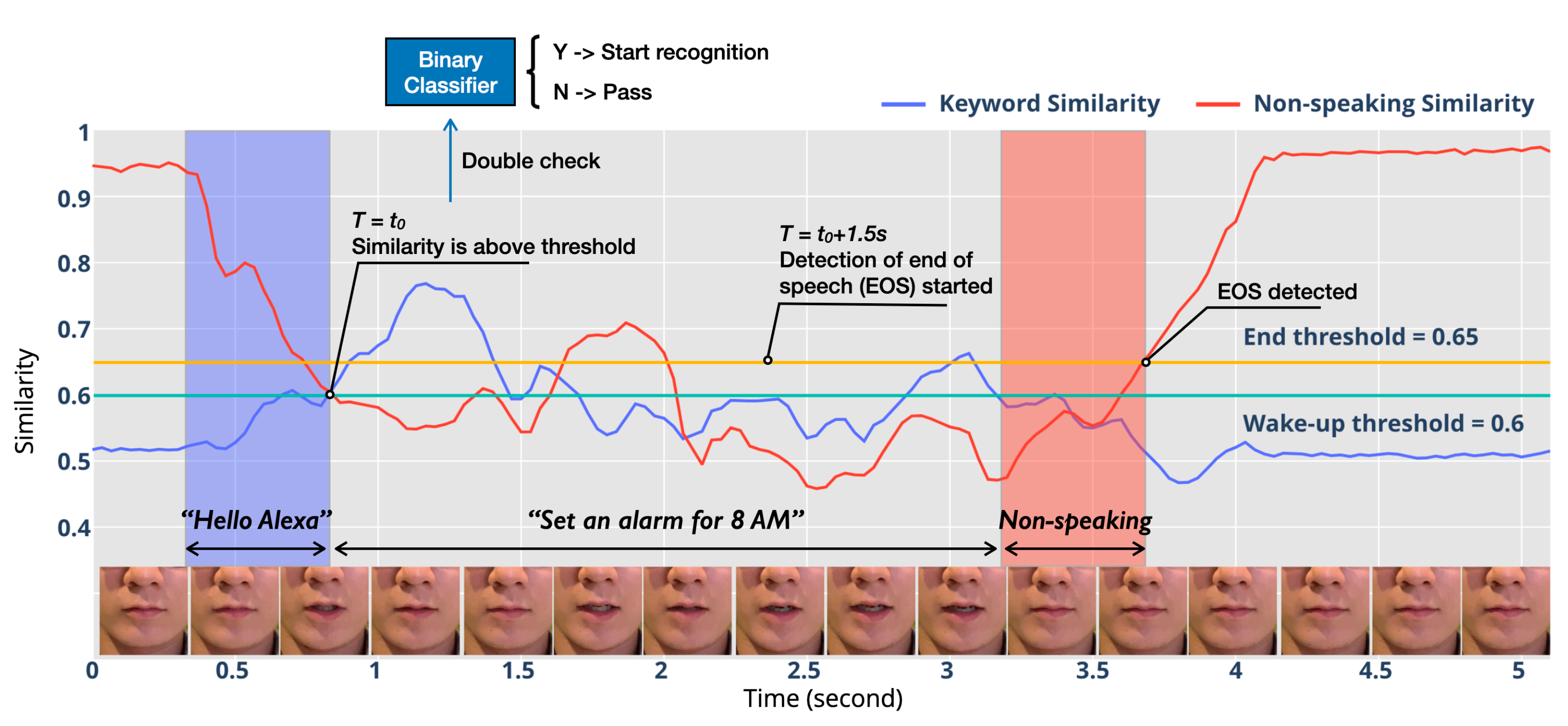}}
    \caption{An example that illustrates the visual keyword spotting technique when the user says "Hello Alexa, set an alarm for 8 AM." When the similarity between the window input and the keyword is above a threshold of 0.6, an additional binary classifier is used to re-examine whether the keyword is issued. If so, the system starts to recognize the following speech as a command. After 1.5s (3 window step size), the system starts to detect the end of the speech (EOS) by calculating the similarity between the window input and the non-speaking sample with a threshold of 0.65.}
    \label{fig:kws}
\end{figure*}
\label{sec:kws}

\subsection{Visual Keyword Spotting}

Detecting and segmenting the user's silent speech has been challenging in real-time lipreading. Previous researchers have proposed to activate the recognition algorithm by using the opening degree of the mouth to identify silent speech~\cite{su2021gaze+, lipinteract, speechin}. However, this approach is prone to misactivation because it can be easily confused when the user is talking to others or unintentionally opens their mouth. 

We propose a few-shot visual keyword spotting method by leveraging the efficient representations extracted by our lip encoder model. Although KWS as an activation method has been predominant in voice interactions, to our best knowledge, this technique has not yet been applied in lipreading-based interfaces. 
Building a KWS model usually requires a huge number of positive and negative training samples, and it is difficult to provide user-defined wake-up keywords. We exploited the generalization ability of the encoder model, which is obtained during the contrastive pre-training, to enable silent keyword detection with customization and rapid calibration. To initialize the KWS module, the user registers a customized phrase as the keyword. Our system then calculates the similarity between the user's real-time lip movements with the keyword utterance sample, (i.e., the cosine similarity between the normalized vectors), thereby spotting when the user is issuing the keyword by comparing the similarity value with a specified threshold. Thus, our technique is available with very few keyword samples and no negative samples.

To determine the optimal threshold for keyword spotting, we leveraged our dataset to estimate the equal error rate (EER) threshold by discriminating one command (deemed as positive samples) over the other commands (deemed as negative samples). The EER results and the corresponding similarity threshold of each command, averaged over all participants, are shown in Figure~\ref{fig:eer}. Overall, our method achieved an average EER of $ 6.75\% $ (standard deviation $2.53\%$). In addition, the number of visemes in the command had a negative correlation with the EER (r=0.688), and an even stronger positive correlation with the EER threshold (r=0.852). This result suggests that using commands with more visemes (i.e., having more complicated lip movements) as the wake-up keyword can yield a lower error rate, but also requires a higher similarity threshold. On the other hand, the optimal threshold can vary widely among individuals. 
For example, for command No.16 \textit{"What time is it"}, the optimal thresholds for P9 and P10 were 0.649 and 0.805. To better understand the data distribution, we visualized the similarity frequency of \textit{Play}, the command with the lowest EER threshold, and \textit{Where's the closest gas station} the command with the highest EER threshold, by using the data from S1 and S2.

Based on these observations, we concluded that although a high keyword spotting accuracy can be achieved using similarity alone, the practical performance optimal threshold can vary considerably depending on the length of the phrase in viseme and the pattern of the user's speech. Therefore, we adopted a relatively low threshold of 0.6, which can accept almost all positive samples over all users and commands in the dataset while still rejecting most negative samples. 
We employed another logistic regression binary classifier to perform a rapid calibration to reduce false positives to discriminate in actual use. As shown in Figure~\ref{fig:interface} C, the user can report when a false positive occurs, and the utterance that has misactivated the system will then be learned as negative. 
Fortunately, as demonstrated in Figure~\ref{fig:kws}, the similarities between non-speaking lip movements were significantly higher, making it much easier to spot the end of the silent speech input.   
Therefore, we only used a similarity threshold of 0.65 without additional classifiers. Furthermore, we set a maximum utterance length of 4s, which means the system will automatically stop recording and perform recognition when the input is longer than 4s.

In real-time use, we used a sliding window of 30 frames (assuming 1s) to extract feature vectors over time. Suspected keyword utterances were detected using the similarity threshold and re-examined using the additional binary classifier. If the utterance is classified as positive, the system is activated and will recognize the subsequent input as a command. Since there is usually a pause between the keyword and the command, the system will start to detect the end of the utterance after a delay of 1.5 times of window length (approximately 1.5s). 


\subsection{System Implementation and Online Incremental Learning Scheme}

We developed an iOS application on an iPhone 13 Pro as a proof-of-concept prototype of LipLearner. The video stream from the front camera was first cropped into the ROIs by using the Vision~\cite{applevision} framework to detect the face and lips. The PyTorch-format lip encoder model was converted into the Core ML~\cite{applecoreml} format, which extracts feature vectors from the ROIs. 
Finally, we employed the MLLogisticRegressionClassifier of the Create ML~\cite{applecreateml} framework to learn the vectors for keyword spotting and silent speech command recognition. The system latency was approximately 250ms feature extraction for 30 frames $+$ 172ms classification $\approx$ 422ms, which is sufficient for real-time interactions. Note that all recognition and fine-tuning processing is done on a commodity mobile phone. Thus, LipLearner can be used without network connections and has all data stored locally, addressing the privacy concerns in lipreading.



Model tests in section~\ref{sec:moreshots} have shown our method can exploit multiple shots for more accurate and robust recognition. To apply this ability in practice, we designed an incremental learning scheme that continuously learns from new data to maximize accuracy (Figure~\ref{fig:interface}). The interaction design of LipLearner can be divided into the following four stages.

\subsubsection{Initialization phase}

To start with, LipLearner will require the user to set up the KWS system for activation and speech segmentation. The user can record several keyword and non-speaking samples by holding the record button at the bottom of the screen. Feature vectors will be extracted from these samples, and the average vectors of each will be used to calculate the similarity for detecting keywords and EOS. As described in Section~\ref{sec:kws}, we also initialized the additional binary classifier with those samples to re-examine suspected keywords. In the subsequent stages, users can report misactivations to improve the KWS classifier.

\subsubsection{Command registration mode}
The user can create novel commands at any time by switching to this mode. To offer a more accessible command registration, we incorporate speech recognition to automatically learn new commands from the voice input using the built-in speech recognizer on iOS 16~\cite{applesfs}. Figure~\ref{fig:interface} B illustrates the registration mode. When the user speaks the new command aloud, LipLearner will record the lip movements and prompt the text recognized from the voice signal as the label. The user can make corrections to the text if incorrect, or just manually input the label if vocalizing is not preferred. Note that to maximize the accuracy, the registration phase also requires the user to first wake up the system using the keyword.  

\subsubsection{Active learning mode}

When the quantity of training data is small (e.g., less than 3 shots), the user can use the system in the active learning mode to improve the model. The system will proactively solicit new data by asking the user to confirm whether the prediction is correct, if not, the user needs to select the correct label from existing commands. Since we only need to re-train the logistic regression classifier part of the model, after new samples are collected, the user can perform on-device fine-tuning at any time. We report that this process can be finished in 2217ms (10-test average) with 30 commands $\times$ 5 shots = 150 samples as training data, suggesting that it is possible to update the model in an in-situ manner.

\subsubsection{On-demand Learning Mode}

If the user thinks that the model has already achieved high performance, they can use the on-demand Learning mode, where the system does not actively collect any data. Instead, the user can choose to correct and add only the misrecognized samples. This mode requires the least effort and prevents the classifier model from overfitting.

\begin{figure}[b]
    \centering
    \makebox[0pt]{\includegraphics[width=0.9\columnwidth]{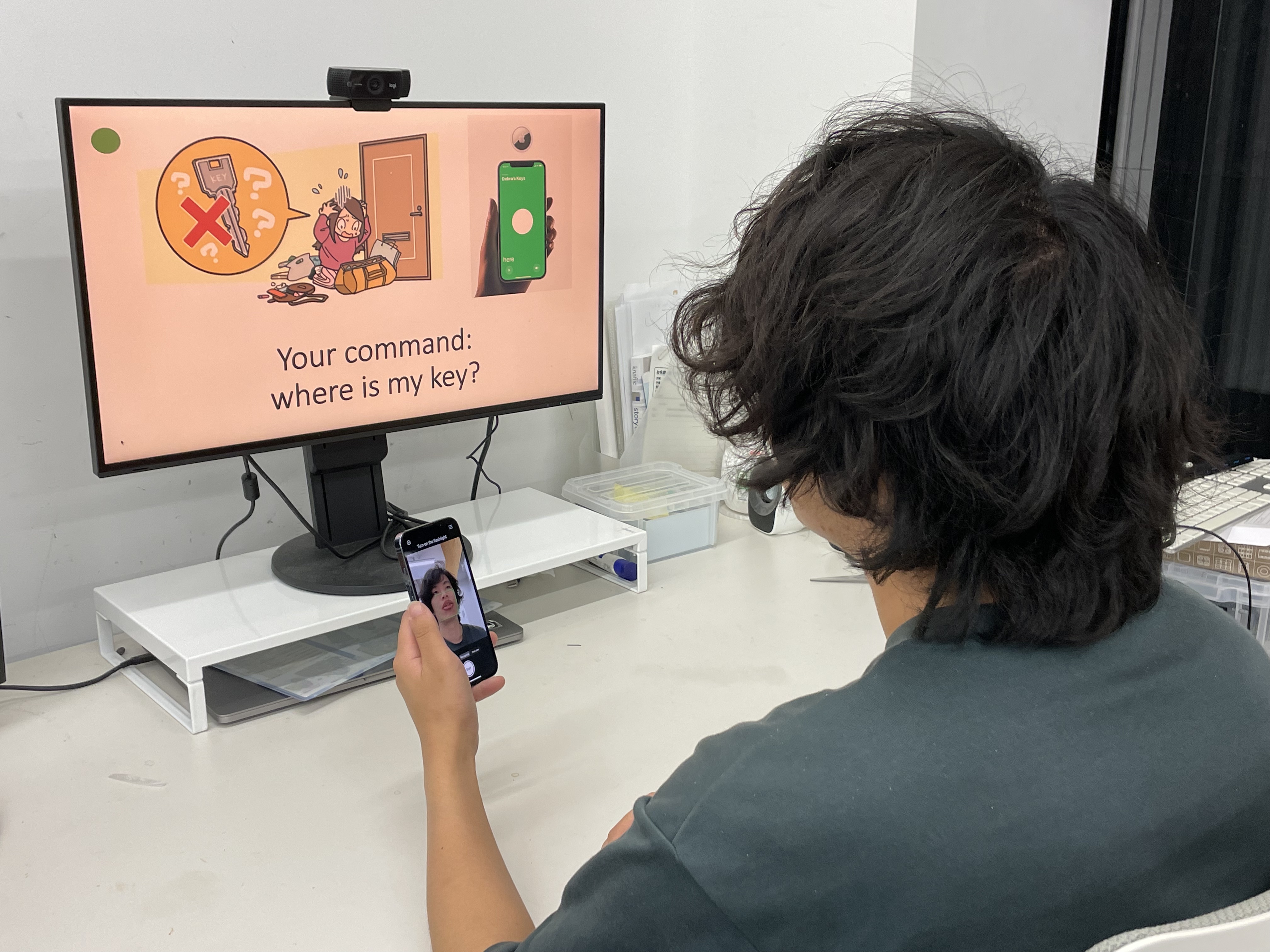}}
    \caption{The user is registering a user-described command that is defined with the guidance of an illustration.}
    \label{fig:user_study}
\end{figure}

\begin{figure*}[t]
    \centering
    \makebox[0pt]{\includegraphics[width=16cm]{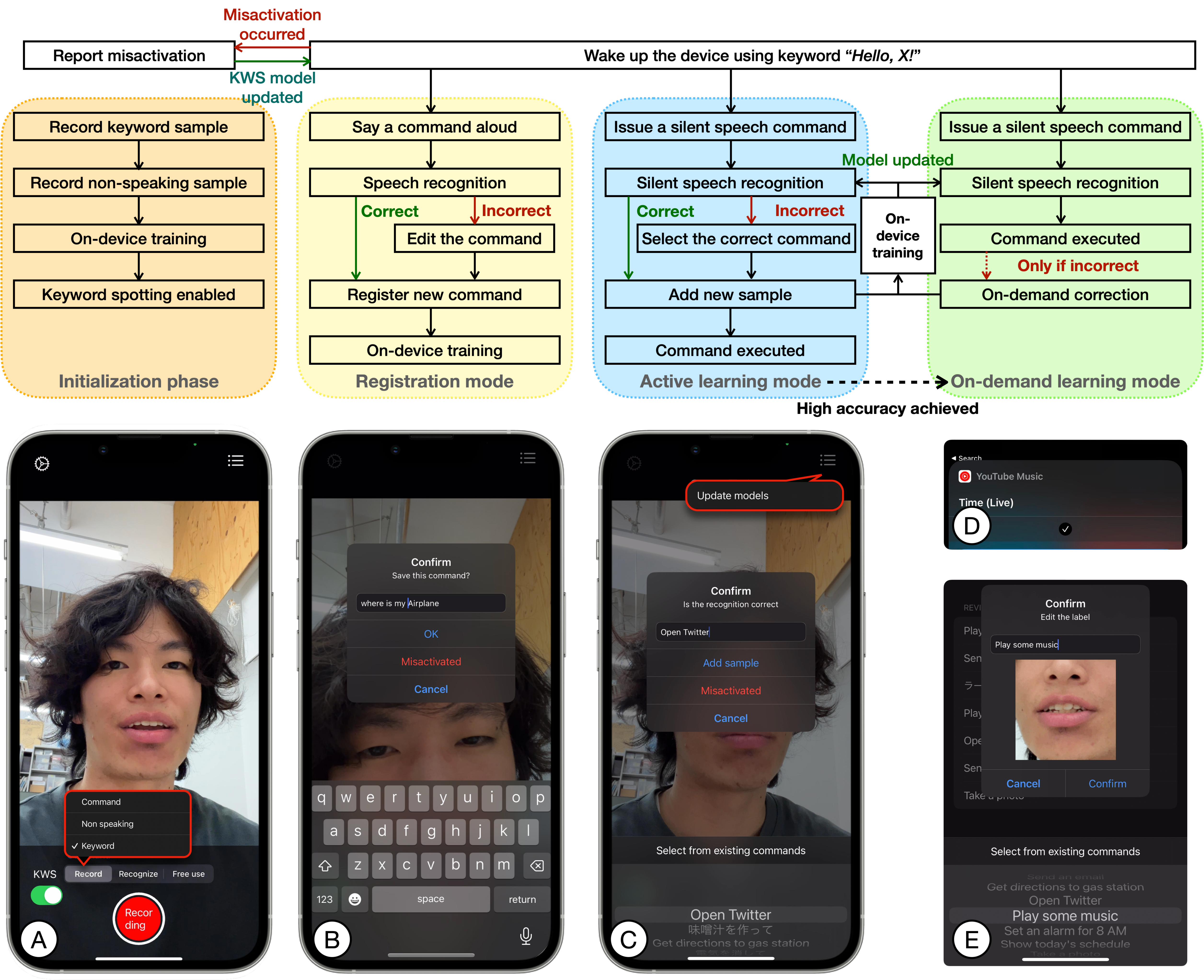}}
    \caption{User experience and interface design. (A) The interface of the initialization phase. The user first needs to record keyword and non-speaking samples to enable KWS activation. (B) The user says a command aloud for command registration. The voice signal will be leveraged to label the silent speech, allowing fast command registration (\textit{Voice2Lip}). (C) The interface for querying the right label in the active learning mode. Users can slide through the existing commands sorted by similarity to select and add a new sample to the model. Users can update the model at any time by using the button at the upper-right corner, which usually takes around 2 seconds on iPhone. (D) An example showing the command "play some music" is recognized correctly and executed successfully by the pre-set shortcut. (E) The interface for correcting the predictions in on-demand learning mode. The user can review recent utterances displayed as a GIF animation.}
    \label{fig:interface}
\end{figure*}

\section{User Study}

We conducted a user study to evaluate LipLearner's usability. This study is distinct from the model test because the silent speech command is issued in real-time and segmented by the KWS algorithm. Furthermore, we wanted to investigate whether our method is able to recognize user-created commands, which can be meant for different intentions with diverse expressions, even in different languages. Finally, it was also important to observe the user's behavior in our human-involved online learning process.

\subsection{Participants and Apparatus}

We recruited 16 participants experienced in using voice assistants to use LipLearner. The participants' native languages are ranging from English, Chinese(including Mandarin, Cantonese, and Hakka), Spanish, Japanese, Malay, and French. This user study also got approved by the university's IRB and all participants were paid 2100 JPY for compensation. 

An iPhone 13 Pro running the LipLearner application was used as the apparatus for the user study. The participants were seated in an armchair and encouraged to hold the phone in the usual way.

\subsection{Design and Procedure}
The user experience design of LipLearner is shown in Figure~\ref{fig:interface} and our user study is consistent with it.


Participants were first given a brief introduction to the system and the interface, after that they were asked to define their wake-up keyword in the format of \textit{"Hello, X"}, where "X" is their preferred name for a smart assistant. Since we have found that phrases with more visemes can provide better KWS performance, "X" was limited to those have more than 3 visemes. Next, participants initialized the system by recording keyword samples and non-speaking samples three times each. Then participants were given five minutes to get themselves familiar with LipLearner by using the activation, command registration, and recognition functions. After participants had sufficiently practiced, they were asked to define their own command set in advance. The command set for user study was divided into three categories based on the level of creative freedom they permit, listed in ascending order as follows:

\begin{enumerate}
\item \textbf{Pre-defined}. We pre-defined 10 English commands (Table~\ref{appendix:fixedcommand} in appendix). Participants were asked to register each command exactly as it is.
\item \textbf{User-described}. We illustrated 10 scenarios where smart assistant could be used (see Figure~\ref{fig:user_study} and Table~\ref{appendix:customcommand}). Participants were asked to use their own words to describe the command they prefer to say in the scenario. There were no restrictions on the language.
\item \textbf{User-created}. Participants were asked to freely create 10 commands with no restrictions or guidance. (Table~\ref{appendix:customcommand}).
\end{enumerate}

Participants registered the 30 commands in one shot using the \textit{Voice2Lip} technique by speaking aloud \textit{"Hello [Name], [Command]"}. Alternatively, they could also choose to input the label manually in cases where they preferred to do so or the speech recognition was not functioning correctly. 

After finishing command registration, participants had a live test session to test LipLearner's performance over six trials. During the test, the experimenter could be directly consulted for clarifications when desired. In each trial, the participant issued each of the 30 commands once. The command to be issued was prompted on a 27-inch monitor in random order. To evaluate the effectiveness of the online incremental learning scheme, the application was set to active learning mode to collect new data from each recognition. If the recognition result was correct, the participant was asked to tap the "add sample" button. Otherwise, they were asked to first select the correct label for the command and then tap the "add sample" button. Upon completion of each trial, LipLearner would obtain a new sample for each command. The participant then could update the model with the new samples by tapping the update button at the top-right corner. In this test, the recognition results were shown on the top of the screen without command execution. We also wanted to verify whether the patterns of lip movements in voiced (normal) speech and silent speech are different, and whether this potential difference would lead to inconsistent recognition performance. To do so, in the first two trials, participants were asked to say the command either in voiced speech or silent speech. The order of the voice trial and the silent trial was counterbalanced among participants. To avoid effects on subsequent trials, only the samples from the silent trial were used for incremental learning as the second shot. 

After the six trials, participants were given 5 to 10 minutes to use LipLearner freely in on-demand learning mode, where they can optionally correct misrecognized commands for better performance. As a proof-of-concept system, we pre-configured the 10 pre-defined commands with the Shortcuts~\cite{appleshortcut} function on iOS, while the other 20 custom commands would still only show the recognition result. We encouraged the participants to experience all pre-configured shortcuts at least once. Finally, they filled out a System Usability Scale (SUS)~\cite{brooke1996sus} questionnaire before attending a semi-structured interview about the experience of using LipLearner.


\section{Results}

\subsection{Observations}
In order to better understand the effect of LipLearner and seek new insights, we noted down the observations during the user study. 

Overall, all participants used the LipLearner smoothly to register and issue silent speech commands. They have personalized LipLearner's names and defined a wide variety of commands (see Table~\ref{appendix:customcommand} in appendix). All non-native English speakers customized the commands in their native language, and 4 participants used more than 2 languages. P12 even used 5 languages to customize commands. This linguistic diversity and promising performance suggest that LipLearner holds the promise of enabling arbitrary language for silent speech. 


In the case of user-defined commands with given scenarios, although some were relatively similar or even the same (e.g. P2, P5, P9 used exactly the same command "\begin{CJK}{UTF8}{gbsn}关灯\end{CJK}" in Chinese), the participants used the expressions that fit their language and speaking habits most. As for the user-created commands, the great richness indicates that LipLearner can exploit much expressiveness of lipreading. 

Some of the participants noticed that LipLearner recognized correctly even if they did not say the commands exactly the same as the commands they have registered. For example, the registered \textit{"what's the weather today"} can be used as \textit{"What's the weather like today"}. The model shows some certain tolerance in all language tested, particularly for minor changes in mid-sentence and end-of-sentence. This nature demonstrates the affinity to real scenarios in which people will register more than 30 commands and may not precisely remember every command. 

\begin{figure}[t]
    \centering
    \makebox[0pt]{\includegraphics[width=0.9\columnwidth]{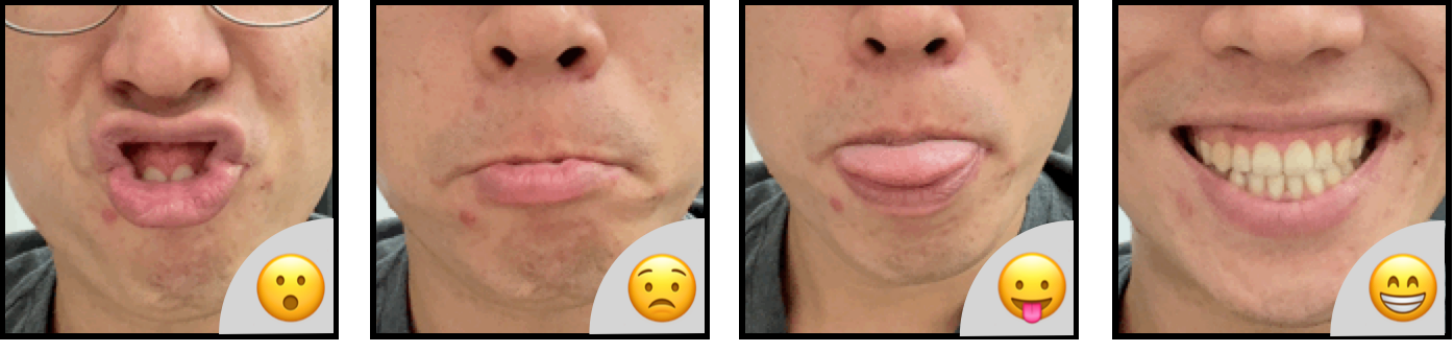}}
    \caption{Facial expression registered as emojis by P10.}
    \label{fig:emoji}
\end{figure}

In the free-use session, P10 tried recording four facial expressions as commands (Figure~\ref{fig:emoji} A) and labeled them with emojis. Since this interesting behavior was never observed before, the experimenter noted down the following recognition results of those expressions. Note that those expressions were recorded in a one-shot manner and classified along with the existing 30 commands. Our system correctly recognized 9 out of 11 tries, and the participant commented, \textit{"It knows what expression I'm trying to make! It's so fun!"} This revealed LipLeaner's potential in recognizing non-verbal commands, which will be discussed later in Section~\ref{sec:non-verbal}. 

\subsection{Quantitative Results}

\begin{figure*}[t]
    \centering
    \makebox[0pt]{\includegraphics[width=14cm]{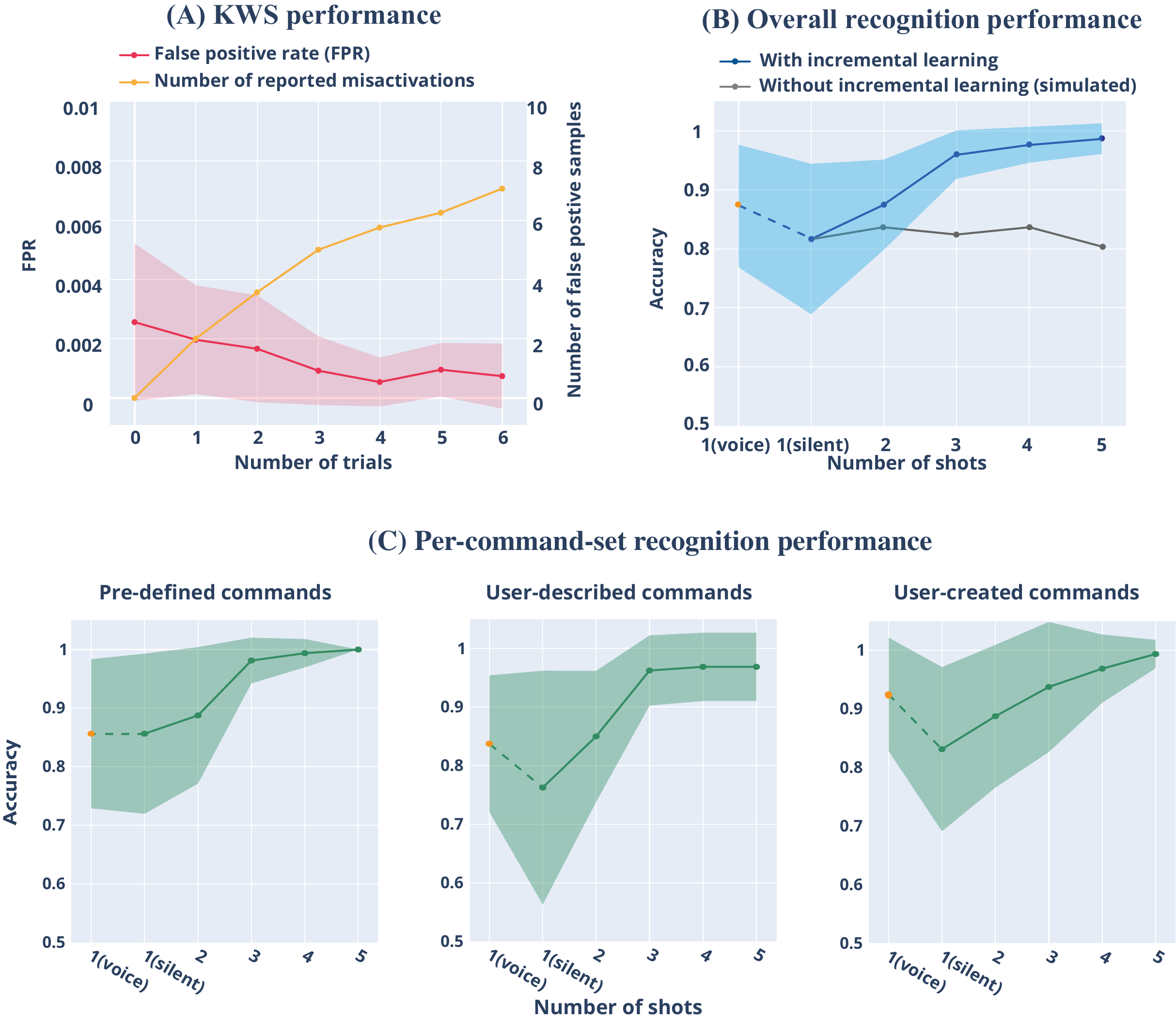}}
    \caption{The false positive rate and recognition performance of LipLearner.}
    \label{fig:user_study_result}
\end{figure*}

\subsubsection{Keyword Spotting performance}

We logged the number of misactivations and false negatives in each trial and depicted it in Figure~\ref{fig:user_study_result} (A). The FPR began from 0.26\% in the first trial and decreased rapidly as the user reported more misactivations, finally achieving 0.07\% with approximately 7 samples. This result indicates that although the KWS function was initialized with only positive samples, it could provide good performance in an early stage and learns efficiently from negative samples over time.

The average false negative rate (FNR) across 7 trials was 1.43\% without notable changes (standard deviation is 0.45\%), because we did not collect positive samples for keywords except in the initialization phase. Note that a lower similarity threshold can reduce false negatives. Although it may also lead to a higher false positive rate (FPR), we think it is admissible given LipLearner's remarkable ability to cope with misactivation. However, since determining the best threshold for all users is impossible, future work should open this setting to the user's choice. 

\subsubsection{Overall Recognition Performance}

As shown in Figure~\ref{fig:user_study_result} (B), first, we find that the one-shot model whose training data all comes from voice input had better accuracy in recognizing vocalized utterances ($87.29\% \pm 10.42\%$) than recognizing unvocalized utterances ($81.67\% \pm 12.80\%$). This suggests that voiced speech and silent speech can have different patterns in lip movements, and learning silent speech from normal speech led to a slight drop in classification accuracy. However, in the post-experiment interview, all participants still expressed a preference for \textit{Voice2Lip} when registering new commands, while using the keyboard to input the command label was considered only when speech recognition fails. Therefore, we believe that sacrificing approximately 5.6\% accuracy in 30-command classification to expedite the command registration process is acceptable.

Furthermore, LipLearner could efficiently expand its knowledge with new samples, which is consistent with the result of the model test. The accuracy rose from $96.04\% \pm 4.12\%$ with 3 shots to $98.75\% \pm 2.60\%$ with 5 shots. Notably, 14 out of 16 participants achieved 100\% accuracy within 5 shots. Most participants favored the on-demand Learning mode because the accuracy was sufficient after finishing the active learning phase and they felt confident using the system ([P7, P9, P15]). To highlight the effect of the online incremental scheme, we simulated a situation where the model did not learn new data during the experiment (Figure~\ref{fig:user_study_result} (B)). We evaluated the system with the same data collected from the user study, while the model was maintained to be the first one-shot model. The result shows that the performance does not improve as the number of trials increases, suggesting that the performance improvement was accomplished solely by incremental learning, instead of the user's familiarization of saying the commands.

\subsubsection{Per Command Set Recognition Performance}

We examined whether LipLearner could provide consistent performance regardless of how the commands were defined by calculating the recognition accuracy in a per-command-set manner~\ref{fig:user_study_result}. In the first silent trial, where the model only used one shot for training, LipLearner achieved better performance on the pre-defined and user-created commands (average accuracy 0.8646 and 0.8500), while the accuracy on user-described commands was lower (average accuracy 0.7646 ). Considering the findings in Section~\ref{sec:kws}, we speculate this difference is caused by the command length. We observed that in the user-described part, participants tended to use short but concise commands to follow the guidance in the illustrations, such as \textit{"Call mom"} and \textit{"Find my car"}. In contrast, user-created commands were longer, more casual, yet full of creativity, e.g., "What are you doing in my swamp!" and "\begin{CJK}{UTF8}{ipxm}さっきの写真をインスタグラムにあげて\end{CJK} (Post my recent photos to Instagram)". The gap among different command sets was closed substantially as more samples were provided. Eventually, all accuracies became above 99\% with 5 shots, demonstrating LipLearner's ability to learn complicated commands in different languages efficiently.

\begin{figure}[t]
    \centering
    \makebox[0pt]{\includegraphics[width=\columnwidth]{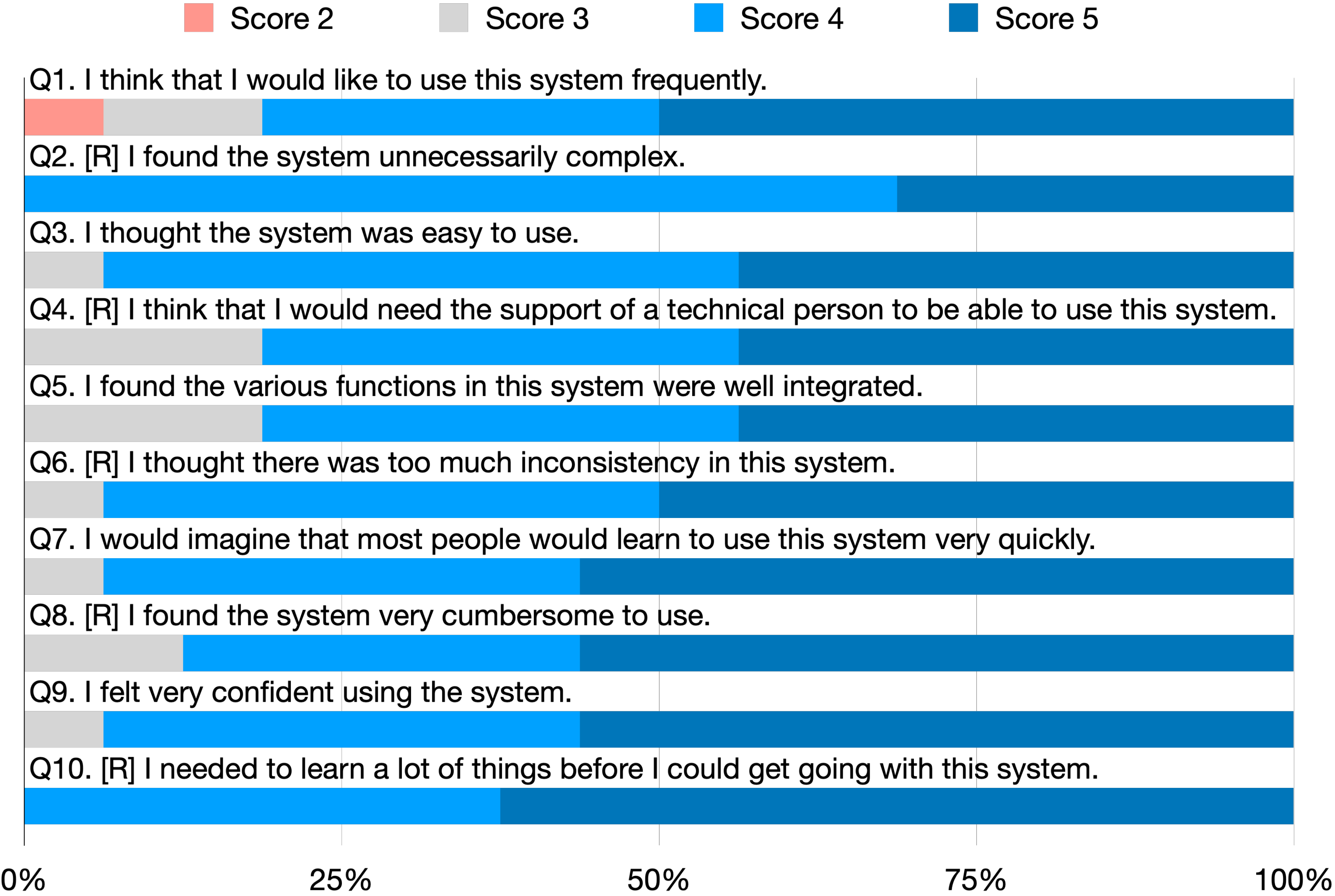}}
    \caption{Usability test results using a 5-scale SUS questionnaire. The horizontal axis is the percentage of responses in each category. Note that the scores of negatively worded statements (Q2,4,6,8,10) are reversed for better visualization.}
    \label{fig:sus}
\end{figure}

\subsection{Qualitative Results}
\subsubsection{Questionnaire Results}
The SUS results suggest generally positive feedback on usability from participants with an overall score of 84.8±6.6, which means it is highly usable and acceptable by users according to Bangor et al.'s empirical evaluation~\cite{bangor2008empirical}. In general, participants expressed confidence in their ability to effectively use the system and rated it as highly easy to use and easy to learn. The details of each SUS question can be found in Figure~\ref{fig:sus}.

\subsubsection{Interview feedback}
We further transcribed the interviews and extracted quotes that were related to user experience and opinions about LipLearner. 
All participants were the first time using a silent speech interface. For the overall usability, 13 out of 16 participants explicitly mentioned that they would like to use LipLearner in the future: \textit{"Now I can use my smart assistant anywhere "}[P2].

Participants were also impressed by the accuracy of the model and the rapid learning process. \textit{"It learns so efficiently, [LipLearner] almost can read all my commands by only listening to me once"}[P9], \textit{"It's amazing that the model can be trained in the blink of an eye."}[P15]

All participants have noticed the improvement in recognition performance, 11 of them found it enjoyable to see the model performs better and better. \textit{"I enjoyed teaching the AI model, it brings me closer to my smart assistant, making it no longer feel like a cold algorithm."}[P7] When asked further how many times they were willing to teach the model, most answers were around 3-5 times. P14 even expressed that \textit{"I am willing to provide more samples for each command since I will gradually enrich my command set instead of immediately registering 30 commands as we did in the user study."}

Some participants further provided suggestions on how we could improve the prototype. Regarding the user interface and interaction, P8 believed that \textit{"The camera view was distracting. I don't think it should necessarily be displayed to users."} and P13 mentioned \textit{"I would be happy if the confirmation process could also be done using silent speech."}

While most of the participants were satisfied with using LipLearner in the on-demand learning mode, P6, P7, and P16 all mentioned about consequences of command execution with misrecognition. \textit{"The commands have different importance and priority. It is better to confirm before the important commands, otherwise, something misrecognized as 'call the police' may lead to a bad consequence."} [P16] 

To conclude, subjective feedback indicated that our system was easy to use and easy to learn, and has provided essential functionalities that allow users to customize their silent speech input experience in real-time.


\section{Discussion}

\subsection{Lipreading Beyond Speech}
\label{sec:non-verbal}
LipLearner benefits from the efficient visual speech representations learned via a contrastive learning strategy. Through our usability studies, we have demonstrated that our method enables to recognize silent speech with a small amount of training data, and its excellent performance can generalize to different phrasing, languages, and even non-verbal lip gestures such as making facial expressions. This ability push forward lipreading beyond speech. 
One potential application is using lipreading for user authentication in complement to face recognition, preventing spoofing attacks and password leakage. The user can define a secret "lip password" by combining several lip gestures, and our few-shot learning technique allows the user to change the password with little effort. Such non-verbal password is difficult to be inferred or remembered by others, therefore being suitable for high-security authentications, e.g., unlocking the device or making a payment. 
Furthermore, although our model is purposed to learn semantic information, we expect the semi-supervised visual speech representations also have the potential to inform user-dependent patterns stemming from subtle lip movements, making it more unlikely to be reproduced by others.
Investigating the difference among individuals can help further understand the feasibility of lipreading-based speaker verification. 

\begin{figure}[b]
    \centering
    \makebox[0pt]{\includegraphics[width=\columnwidth]{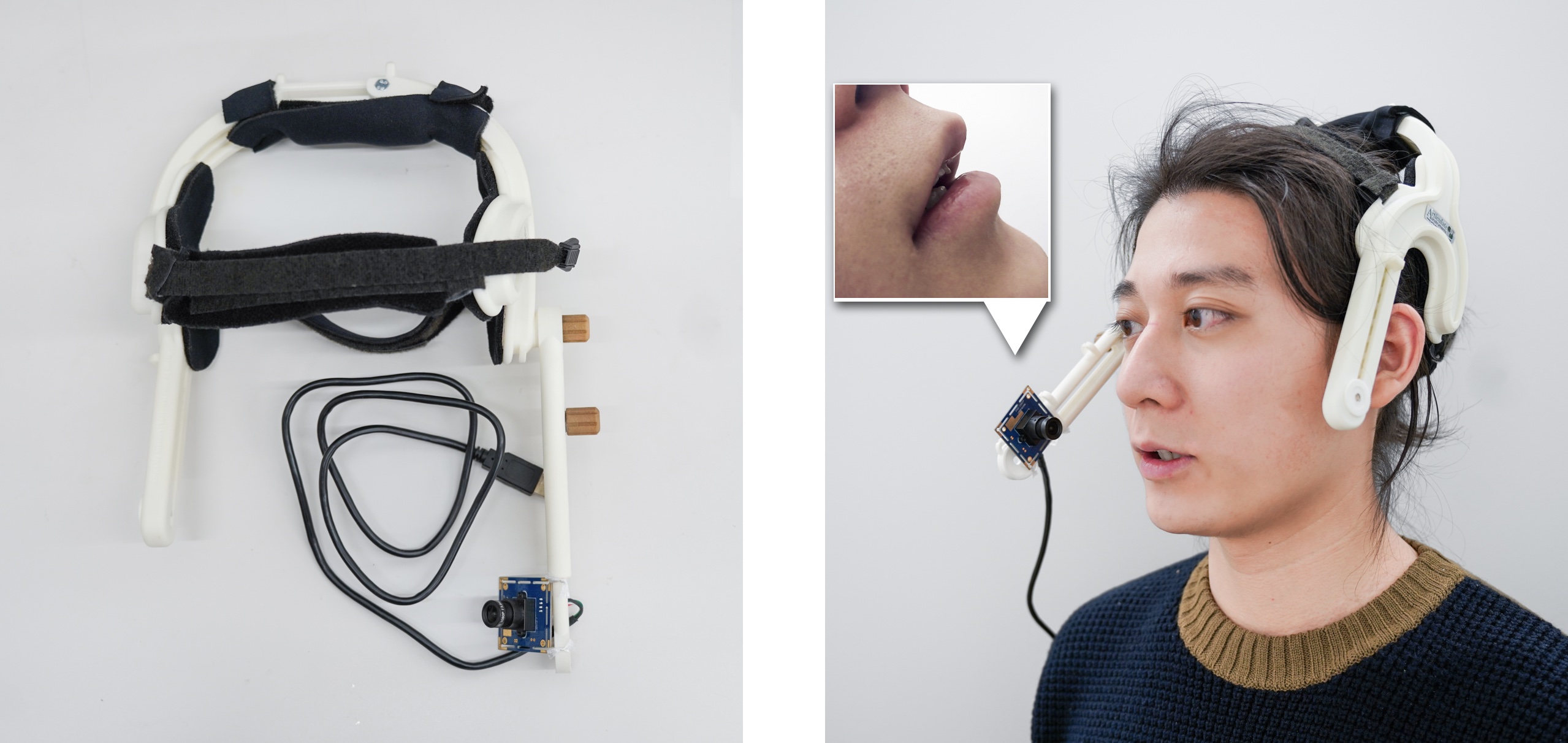}}
    \caption{The device used for the preliminary test on wearable lipreading using our few-shot customization framework.}
    \label{fig:wearable}
\end{figure}

\subsection{Towards Wearable Lipreading}

This research is based on mobile interactions because of the prevalence of smartphones. However, we believe lipreading technologies can facilitate communication between humans and computers in a diversity of scenarios. The recent boom in head-mounted displays (HMD) based VR/AR applications calls for natural input methods with high mobility. 
Lipreading is a promising approach for its expressiveness and low learning cost, and it can be easily implemented by embedding a lip-observing camera in the headset. However, lipreading at such a close distance is not trivial because capturing the mouth usually requires a fish eye camera, whose distortion effects can pose challenges for recognition. Yet, placing the camera in the front of face is obtrusive. Our method in contrast has shown a consistently good performance recognizing from different points of view. To explore the feasibility of applying LipLearner in wearable scenarios, we did a preliminary study by mounting a USB camera on a 3D-printed headset (Figure~\ref{fig:wearable}) that captures the user's profile face. We collect a dataset from one of the authors with the same command set used in Section~\ref{sec:data_collection}, making up a dataset of 25 commands $\times$ 4 repetitions = 100 samples. We evaluated the system's performance by running an offline test on a PC, and the 1-shot, 2-shot, and 3-shot accuracies are 0.7941, 0.9387, 1.0 (averaged over 100 random seeds). These early results indicate that our model can achieve good performance even recognizing profile faces. Furthermore, the visual KWS technique can free users' hands and better make them immersed in the virtual worlds. This preliminary study demonstrates that our few-shot lipreading framework holds the promise of extending the dimensions of VR/AR interactions.


\subsection{Human-in-the-loop Incremental Learning}

LipLearner sheds new light on human-in-the-loop interactions by focusing on offering a natural and easy way to involve users. Instead of immediately requiring enormous data to pursue high accuracy, we introduce a one-shot command registration technique \textit{Voice2Lip} to allow rapid initialization. LipLearner proactively solicits new samples from the user when the data is insufficient, and learns in an on-demand mode when high accuracy is achieved. Feedback from the user study suggested that participants enjoyed this human-AI interaction, and they were willing to help with improving the AI system during use. We envision that in the future, the design space of how to engage users to provide knowledge for learnable AI systems, such as minimizing the disruptions, will be an important topic in HCI.

\section{Limitations and Future Work}

While LipLearner demonstrates favorable usability, there are several key limitations that will need to be overcome in the future. 

First, there is still room to lessen the physical and cognitive labor of active learning. Several participants mentioned despite the fact that they enjoyed helping improve the model in the active learning mode, it would be better to be able to validate or correct the predictions also using silent speech (e.g., saying \textit{"Yes"} or \textit{"Cancel"}) instead of tapping buttons. Although this feedback also indicates that silent speech is preferred for its low effort in mobile interactions, the interaction design should be optimized to better involve the user in the human-in-the-loop flow. 

Second, although our user study observations revealed LipLearner's tolerance for minor changes of expressions, this may make it more difficult to distinguish very similar commands. For example, we find that one of the common misrecognition is between \textit{"Turn on the light"} and \textit{"Turn on the flashlight"}. The problem can be alleviated by proactively soliciting more samples for low-accuracy commands or asking the user to rephrase. 

Undoubtedly, few-shot learning has enhanced silent speech by extending the vocabulary capacity and minimizing the user burden in command registration. However, due to the lack of context, the level of abstraction of lip commands is still relatively low. For example, two separate commands need to be registered to set the alarm for 8 AM and 9 AM. We envision that the expressiveness and abstraction level of LipLearner can be further boosted by training zero-shot lipreading models jointly with language models such as GPT3~\cite{NEURIPS2020_1457c0d6} or T5 ~\cite{2020t5}. In zero-shot lipreading, the user only has to prepare a bunch of command candidates they would like to use, and the model can recognize completely unseen commands by matching lipreading embeddings with text embeddings.

\section{Conclusion}

This paper presents LipLearner, a lipreading-based silent speech interface that enables in-situ command customization on mobile devices. We leverage contrastive learning to build a model to learn efficient visual speech representations from public datasets, providing in-situ fine-tuning for unseen users and words using few-shot learning. For a preliminary test, we collected a dataset covering various mobile interaction scenarios to evaluate the model's performance and robustness against lighting conditions, user posture, and hold gestures. The result showed that our method could provide consistent performance in different settings, outperforming conventional supervised methods. To investigate usability, we developed a prototype of LipLearner on iOS by integrating the few-shot customization framework with an online incremental learning scheme, involving the user in the learning process to improve the model on their demand. We further minimize the labor of command registration and incorporate speech recognition to automatically learn new commands from voice input. Through a user study, we demonstrated that LipLearner also has excellent performance with various commands defined by participants in different languages. The subjective feedback suggested that LipLearner is easy to use and easy to learn, and most participants enjoyed the human-AI integrated interaction. \zxsucr{To conclude, our system democratizes silent speech by offering quick-start on-device lipreading, and it unleashes users' creativity with customizable commands. We hope our work can bring the vision of human-centered AI closer to reality, spotlighting the importance of intuitive and personalized interaction experiences.}


\begin{acks}
 This work was supported by JST Moonshot R\&D Grant Number JPMJMS2012, JST CREST Grant Number JPMJCR17A3, The University of Tokyo Human Augmentation Research Initiative, and a collaborative research fund between Mercari Inc. R4D and RIISE. We would further like to thank the anonymous reviewers for their constructive feedback and the participants of the user study.
\end{acks}

\bibliographystyle{ACM-Reference-Format}
\bibliography{manuscript}

\appendix

\clearpage
\onecolumn
\section{User Study Command Set} \label{userstudycommandset}

\begin{table}[h]
    \centering
    \small
    \caption{10 Pre-defined Commands.}
    \label{appendix:fixedcommand}
    \begin{tabular}{ll}
    Get directions to gas station & Take a photo             \\
    Open Twitter                  & Turn on focus mode       \\
    Play some music               & Turn on the flashlight   \\
    Send an email                 & What’s the weather today \\
    Set an alarm for 8 am         & Show today’s schedule   
    \end{tabular}
\end{table}

\begin{CJK*}{UTF8}{gbsn}
\tiny
\begin{center}
    
\begin{longtable}{p{0.1\textwidth}p{0.2\textwidth}p{0.2\textwidth}p{0.2\textwidth}p{0.2\textwidth}}
\caption{20 custom commands. B1 to B10 are the user-described commands with given scenarios. C1 to C10 are the user-created commands. \label{appendix:customcommand}}\\

\multicolumn{1}{l}{\textbf{Participant}}   & \textbf{P1}                               & \textbf{P2}                            & \textbf{P3}                                   & \textbf{P4}                                        \\
\multicolumn{1}{l}{\textbf{Languages}} & \textbf{French, Japanese, English}        & \textbf{Chinese, English}              & \textbf{Chinese, Japanese, English}           & \textbf{Spanish, English}                          \\
\multicolumn{1}{l}{\textbf{Keyword}}       & \textbf{Hello, Mirai}         & \textbf{Hello, Baymax}              & \textbf{Hello, Mugi}           & \textbf{Hello, David}  \\
B1                                & \begin{CJK}{UTF8}{ipxm}キーはどこ\end{CJK}  & 我钥匙呢                                & Where is my key                               & Donde estan mis llaves                             \\
B2                                & Apelle maman                              & 打给妈妈                                & Call my mom                                   & Llama a mama                                       \\
B3                                & Ouvre les rideaux                         & 拉开窗帘                                & Open the curtain                              & Abre las cortinas                                  \\
B4                                & Reserve un ticket d’avion pour le Japon   & 去东京的机票多少钱                       & Buy the ticket to tokyo                        & Compra pasaje de vuelo                             \\
B5                                & \begin{CJK}{UTF8}{ipxm}電気消して\end{CJK}  & 关灯                                   & Turn off the light                            & Prende las luces                                   \\
B6                                & Order food                                & 最近的汉堡店在哪                         & Reserve a restaurant                          & Llama a ubereats                                   \\
B7                                & \begin{CJK}{UTF8}{ipxm}最近のニュースはどう\end{CJK}   & 今天有什么新闻                     & Open Japan today                              & Dime las noticias de hoy                           \\
B8                                & 6 minutes timer                           & 倒计时6分钟                                 & Count six minutes                             & Pon cronometro de seis minutos                     \\
B9                                & Where is my car                           & 我车呢                                    & Find my car                                   & Encuentra mi auto                                  \\
B10                               & Monte le chauffage                        & 关空调                                    & Turn off the air conditioner                  & Prende la calefaccion                              \\
C1                                & Read this                                 & 带我去最近的超市                            & \begin{CJK}{UTF8}{ipxm}分かりません\end{CJK}    & Reinicíate                                         \\
C2                                & \begin{CJK}{UTF8}{ipxm}一番近いゲーセン\end{CJK}  & 现在几点                            & \begin{CJK}{UTF8}{ipxm}また来週\end{CJK}        & Abre traductor de google                           \\
C3                                & Comment ca va?                            & 明天八点叫我起床                             & \begin{CJK}{UTF8}{ipxm}だめだね\end{CJK}       & Call father                                        \\
C4                                & I am having a lot of fun                  & 东京大学怎么样                                & Do some research                              & Edit photo                                         \\
C5                                & Quand ouvre le cinema                     & 播放《武林外传》                               & 关闭声音                                          & Dim screen to minimun                              \\
C6                                & What are you doing in my swamp            & 打开亚马逊搜索                                & 给我点钱                                          & Delete photo                                       \\
C7                                & Name all the pokemons                     & 帮我打的去机场                                & 帮我做饭                                          & Lock my screen                                     \\
C8                                & Chante une chanson                        & 提醒我下周三上午九点有会议                          & 打电话给爸爸                                        & Open clash of clans                                \\
C9                                & Create a macro for my lilly heals         & 打开日语翻译器                                & 打开电视机                                         & Download file                                      \\
C10                               & When is the next HatsuneMiku concert      & 豆沙馅怎么做                                 & 放点轻松音乐                                        & Share video                                        \\
                                  &                                           &                                        &                                               &                                                    \\
 \multicolumn{1}{l}{\textbf{Participant}}    & \textbf{P5}                               & \textbf{P6}                            & \textbf{P7}                                   & \textbf{P8}                                        \\
\multicolumn{1}{l}{\textbf{Language Used}}  & \textbf{Chinese, English}                 & \textbf{English}                       & \textbf{English}                              & \textbf{Japanese, English}                         \\
\multicolumn{1}{l}{\textbf{Keyword}}       & \textbf{Hello, Mamun}         & \textbf{Hello, Mr. Ha}              & \textbf{Hello, friend}           & \textbf{Hello, David}  \\
B1                                & Where’s my key                            & Can you find my key                    & Where’s my key                                & \begin{CJK}{UTF8}{ipxm}鍵をさがして\end{CJK}                     \\
B2                                & Call mom                                  & Call Mom                               & Call my mom                                   & \begin{CJK}{UTF8}{ipxm}お母さんをよんで\end{CJK}                   \\
B3                                & Open curtains                             & Open the curtain                       & Open curtain                                  & \begin{CJK}{UTF8}{ipxm}カーテンをあけて\end{CJK}                   \\
B4                                & 预定飞机票                                     & I want to book a ticket to Tokyo       & Find me a ticket to tokyo                     & \begin{CJK}{UTF8}{ipxm}航空券を予約して\end{CJK}                \\
B5                                & 关灯                                        & Turn to sleep mode                     & Turn off light                                & \begin{CJK}{UTF8}{ipxm}電気を消して\end{CJK}                      \\
B6                                & 预定餐厅                                      & Find the nearest restaurant            & Find me some restaurants                      & \begin{CJK}{UTF8}{ipxm}レストランを予約して\end{CJK}              \\
B7                                & 浏览新闻                                      & What’s the news today                  & Read me some news                             & \begin{CJK}{UTF8}{ipxm}ニュースを読んで\end{CJK}                  \\
B8                                & 设定6分钟的计时器                                 & Set a timer for 6 minutes              & Set a 6 minutes timer                         & \begin{CJK}{UTF8}{ipxm}６分間のタイマーをセットして\end{CJK}      \\
B9                                & 寻找我的车                                     & Find my car                            & Where did I park my car                       & \begin{CJK}{UTF8}{ipxm}車を探して\end{CJK}                        \\
B10                               & 打开暖风                                      & Turn on the air conditioner            & Turn off the aircon                           & \begin{CJK}{UTF8}{ipxm}暖房をつけて\end{CJK}                       \\
C1                                & 整理相册                                      & Tell me how to say ”sorry” in Japanese & How to go to my university                    & \begin{CJK}{UTF8}{ipxm}一ドルは何円\end{CJK}                        \\
C2                                & 打开投影仪                                     & Am I smart?                            & Text my mom                                   & \begin{CJK}{UTF8}{ipxm}今日の予定を教えて\end{CJK}                  \\
C3                                & 设定闹钟                                      & Delete Wechat                          & Clean my house                                & \begin{CJK}{UTF8}{ipxm}今日の天気を教えて \end{CJK}                  \\
C4                                & 今天的股票价格                                   & Buy some pork in Amazon                & Play happy eliminating                        & \begin{CJK}{UTF8}{ipxm}フェイスブックを開けて\end{CJK}              \\
C5                                & 你的工作是什么                                   & Update my calendar                     & Show my calendar                              & \begin{CJK}{UTF8}{ipxm}自宅までの距離は？\end{CJK}                  \\
C6                                & 播放音乐                                      & Where is the bus stop                  & Where’s the nearest hospital                  & \begin{CJK}{UTF8}{ipxm}クラシック音楽をかけて\end{CJK}                 \\
C7                                & 讲个笑话吧                                     & Call uber to my hometown               & USD to Japanese yen                           & \begin{CJK}{UTF8}{ipxm}ドアを開けて\end{CJK}                         \\
C8                                & 最近的音乐会有哪些                                 & Install Google                         & What’s gravity                                & \begin{CJK}{UTF8}{ipxm}大阪までの経路を教えて\end{CJK}                    \\
C9                                & 今天的天气如何                                   & Buy 1 million stocks                   & Open camera                                   & \begin{CJK}{UTF8}{ipxm}明日の７時に締め切りをリマインドして\end{CJK}        \\
C10                               & 导航回家                                      & Call my lover                          & Turn off volume                               & \begin{CJK}{UTF8}{ipxm}卵焼きの作り方を教えて\end{CJK}                 \\
                                  &                                           &                                        &                                               &                                                    \\
                                 \\
                                \\
                                \\
                                \\
                                \\
                                \\
\multicolumn{1}{l}{\textbf{Participant}}   & \textbf{P9}                               & \textbf{P10}                           & \textbf{P11}                                  & \textbf{P12}                                       \\
\multicolumn{1}{l}{\textbf{Language used}}   & \textbf{Chinese, English}                 & \textbf{Chinese,English}              & \textbf{Chinese, English}                     & \textbf{Chinese(Cantonese, Hakka), Malay, Japanese, English} \\
\multicolumn{1}{l}{\textbf{Keyword}}       & \textbf{Hello, Tom}         & \textbf{Hello, Alexa}              & \textbf{Hello, Jessica}           & \textbf{Hello, David}  \\
B1                                & 我的钥匙在哪                                    &  Find my key                            & Where is my key                               & 我的钥匙叻                                              \\
B2                                & 打电话给妈妈                                    & Make a phone call to Mom               & Make a call                                   & Call mami                                          \\
B3                                & 打开窗帘                                      & Open the curtain                       & Open the curtain                              & Wake me up at 9                                    \\
B4                                & 帮我买去东京的机票                                 & Book a flight ticket                   & Book a flight ticket                          & Tolong beli tiket ke jepun                         \\
B5                                & 关灯                                        & Turn on the light                      & Turn off light                                & Nak tidur ni                                       \\
B6                                & 查找附近的餐厅                                   & Make a reservation of restaurants      & Find a restaurant                             & \begin{CJK}{UTF8}{ipxm}いい感じのレストラン探して\end{CJK}                                      \\
B7                                & 打开新闻                                      & What’s today’s news                    & Show news                                     & Pull up today’s news                               \\
B8                                & 倒计时六分钟                                    & Set a timer for six minutes            & Set the alarm clock at 6                      & Set the timer to 6                                 \\
B9                                & 我的车在哪                                     & where is my car                        & Find my car                                   & \begin{CJK}{UTF8}{ipxm}車どこ\end{CJK}                                                \\
B10                               & 提高空调的温度                                   & Turn on the air conditioner            & Turn up the temperature                       & 太冷了                                                \\
C1                                & 提高耳机音量                                    & 放几首歌听                                  & Tell a joke                                   & I want to be rich                                  \\
C2                                & 帮我回复妈妈的微信，好的                              & 今天天气好吗                                 & Turn off camera and microphone                & I will be back                                     \\
C3                                & 早上8点开始洗衣服                                 & 设定一个明早九点的闹钟                            & Clean the trash bin                           & Esok jangan lupa BBQ ye                            \\
C4                                & 打扫房间                                      & 今天天气怎么样                                & How old are you?                              & \begin{CJK}{UTF8}{ipxm}きみ、かわいいね\end{CJK}               \\
C5                                & 今天天气怎么样                                   & 怎样去学校                                  & Navigate me to the conference center          & \begin{CJK}{UTF8}{ipxm}あらあら\end{CJK}                    \\
C6                                & 明天9点叫我起床                                  & 明天会下雨吗                                 & Calculate twenty three hundred divided by six & \begin{CJK}{UTF8}{ipxm}ラーメンは煮干しでしょ\end{CJK}    \\
C7                                & 我的手机在哪                                    & 打电话给小李                                 & 推荐一些新书                                        & \begin{CJK}{UTF8}{ipxm}CHI論文通して\end{CJK}                                         \\
C8                                & 提醒我明天交作业                                  & 附近有什么餐厅                                & 叫个外卖                                          & 可以点菜了吗                                             \\
C9                                & 每小时帮我倒水                                   & 放一首周杰伦                                 & 未来一周的天气如何                                     & 唔中意日本                                              \\
C10                               & 附近的商场有哪些                                  & 帮我发条短信                                 & 新建文件夹                                         & 这个\begin{CJK}{UTF8}{ipxm}たこ焼き\end{CJK}不错                                           \\
                                  &                                           &                                        &                                               &                                                    \\
\multicolumn{1}{l}{\textbf{Participant}}  & \textbf{P13}                              & \textbf{P14}                           & \textbf{P15}                                  & \textbf{P16}                                       \\
\multicolumn{1}{l}{\textbf{Language used}}  & \textbf{Japanese, English}                & \textbf{Japanese, English}             & \textbf{Chinese, Japanese, English}           & \textbf{English}                                   \\
\multicolumn{1}{l}{\textbf{Keyword}}       & \textbf{Hello, Alexa}         & \textbf{Hello, Alexa}              & \textbf{Hello, Oliver}           & \textbf{Hello, Thomas}  \\
B1                                & Where is my key?                          & I lost my key                          & 钥匙去哪了                                         & Looking for my key                                 \\
B2                                & Hello mom                                 & Call my mom                            & 给妈妈打电话                                        & Call mom                                           \\
B3                                & Good morning                              & \begin{CJK}{UTF8}{ipxm}カーテンを開けて\end{CJK}                               & 打开窗帘                                          & Draw the curtains                                  \\
B4                                & \begin{CJK}{UTF8}{ipxm}一番安い飛行機はどれ？\end{CJK}                               & I want to go to tokyo                  & 查询一下去东京的机票                                    & Check for tickets to tokyo                         \\
B5                                & \begin{CJK}{UTF8}{ipxm}電気を消して\end{CJK}                                    & Turn off the light                     & 关灯                                            & Turn off the lights                                \\
B6                                & \begin{CJK}{UTF8}{ipxm}一番近いレストランはどこ\end{CJK}                              & \begin{CJK}{UTF8}{ipxm}お腹すいた\end{CJK}                                  & 有没有什么推荐的餐厅                                    & Find a restaurant for me                           \\
B7                                & \begin{CJK}{UTF8}{ipxm}ニュースを開いて\end{CJK}                                  & open news app                          & 今天有什么新闻                                       & What’s news today                                  \\
B8                                & Set Timer                                 & Set a timer                            & 倒计时六分钟                                        & Countdown 6 minutes                                \\
B9                                & where is my car                           & Find my car                            & 我的车子在哪                                        & Where did I park my car                            \\
B10                               & \begin{CJK}{UTF8}{ipxm}エアコンの温度を下げて\end{CJK}                               & \begin{CJK}{UTF8}{ipxm}暖房つけて\end{CJK}                                  & 调高空调温度                                        & Warm up here                                       \\
C1                                & Is Singularity already here?              & \begin{CJK}{UTF8}{ipxm}お腹いっぱいです\end{CJK}                               & \begin{CJK}{UTF8}{ipxm}家への経路を教えて\end{CJK}                                     & Check formula 1 schedule                           \\
C2                                & Make collage lunches better               & \begin{CJK}{UTF8}{ipxm}家に帰りたい\end{CJK}                                 & 明早7点叫我起床                                      & USD to Japanese yen                                \\
C3                                & Write a book that sells well              & \begin{CJK}{UTF8}{ipxm}喉が渇いた\end{CJK}                                  & \begin{CJK}{UTF8}{ipxm}さっきの写真をインスタグラムにあげて\end{CJK}                            & Check youtube updates                              \\
C4                                & What is the raw material of these clothes & \begin{CJK}{UTF8}{ipxm}電源を消して\end{CJK}                                 & \begin{CJK}{UTF8}{ipxm}ラーメン食べたい\end{CJK}                                      & Monitoring my dog at home                          \\
C5                                & Grow houseplants                          & Find my laptop                         & 麻婆豆腐怎么做                                       & Next month’s bills                                 \\
C6                                & \begin{CJK}{UTF8}{ipxm}味噌汁を作って\end{CJK}                                   & Misactivate                            & \begin{CJK}{UTF8}{ipxm}電気を消して\end{CJK}                                        & Play my daily mix                                  \\
C7                                & \begin{CJK}{UTF8}{ipxm}ハンバーガーのハンバーグ抜きを頼んで\end{CJK}                        & Tell me a joke                         & \begin{CJK}{UTF8}{ipxm}静かにして\end{CJK}                                         & Check out the nearby exhibitions                   \\
C8                                & \begin{CJK}{UTF8}{ipxm}遅刻の言い訳を考えて\end{CJK}                                & Say hello                              & \begin{CJK}{UTF8}{ipxm}今日の終電は何時 \end{CJK}                                     & Call the police                                    \\
C9                                & \begin{CJK}{UTF8}{ipxm}日本メタバース協会ってなに\end{CJK}                             & What time is it                        & \begin{CJK}{UTF8}{ipxm}最近のヒット曲を再生して\end{CJK}                                  & Todo list tomorrow                                \\
C10                               & \begin{CJK}{UTF8}{ipxm}私の博士論文を書いて\end{CJK}                                & What’s your name                       & \begin{CJK}{UTF8}{ipxm}車借りて\end{CJK}                                          & Open Netflix                                      
\end{longtable}

\end{center}
\end{CJK*}

\end{document}